\newcommand{\be}{\begin{equation}}
\newcommand{\ee}{\end{equation}}
\newcommand{\bea}{\begin{eqnarray}}
\newcommand{\eea}{\end{eqnarray}}
\documentclass[epjc3,nopacs]{svjour3}
\pdfoutput=1
\usepackage[T1]{fontenc}
\usepackage{graphicx}
\usepackage{dcolumn}
\usepackage{bm}
\usepackage{amsmath}
\usepackage{amssymb}
\def\d{d\kern-.8 ex\vrule height 1.3 ex depth-1.24 ex width .7 ex \kern .15 ex}
\def\D{D\kern-1.7 ex\vrule height .87 ex depth-.8 ex width .7 ex \kern .95 ex}

\journalname{Eur. Phys. J. C}

\begin{document}

\title{Wormholes and out-of-time ordered correlators in gauge/gravity duality}

\author{Mihailo \v{C}ubrovi\'c\thanksref{e1,adr1}}
\thankstext{e1}{e-mail: cubrovic@ipb.ac.rs}
\institute{Center for the Study of Complex Systems, Institute of Physics Belgrade, University of Belgrade, Pregrevica 118, 11080 Belgrade, Serbia\label{adr1}}

\date{\today}
\maketitle

\abstract{
We calculate the four-wave scattering amplitude in the background of an AdS traversable wormhole in 2+1 dimensions created by a nonlocal coupling of AdS boundaries in the BTZ black hole background. The holographic dual of this setup is a pair of CFTs coupled via a double-trace deformation, the scattering amplitude giving the out-of-time ordered correlator (OTOC) in CFT. Short-living wormholes always exhibit a regime of fast scrambling, saturating the MSS bound for the Lyapunov exponent but in the early-time regime the scrambling can be slower, with a Lyapunov exponent linear in $T$ but below the MSS bound. For long-living (near-eternal) wormholes however the numerics suggests the existence of another regime, with a drastic (exponential) slowdown of scrambling and an exponentially small Lyapunov exponent. Our findings have parallels in the SYK model, and may indicate certain limitations of wormhole teleportation protocols previously studied in the literature.
}

\section{Introduction}\label{sec1}

For a long time wormhole geometries have held the status of an intriguing curiosity but little more than that. Quite early on, it was shown \cite{thorneamjp,visser} that a wormhole requires the violation of even the weakest energy condition, the null energy condition (NEC). To do that, one needs either nonclassical matter or exotic, non-minimally coupled classical matter such as conformally coupled scalars or nonlocal interactions. Physically motivated examples of such interactions have been found to give traversable wormholes in asymptotically flat space, with help of Landau-quantized fermions in magnetic field \cite{worm4} or cosmic strings \cite{wormmarstring}, and in AdS space, through nonlocal coupling via a double-trace deformation \cite{wormjaff,wormbound,wormhighd,wormads4,wormfrei} or via a quotient by a discrete isommetry \cite{wormmar1,wormmar2}, or again through cosmic strings \cite{wormhor}. For AdS/CFT, the 2+1-dimensional wormhole obtained by Gao, Jafferis and Wall (GJW) from a double-trace deformation of the BTZ black hole \cite{wormjaff} is a particularly natural setup: being asymptotically AdS, it has a field theory dual, and living in 2+1 dimensions the dual is a proper (1+1-dimensional) quantum field theory rather than quantum mechanics (i.e., 0+1-dimensional theory). Higher-dimensional analogue
of the GJW protocol is also possible and has been constructed in \cite{wormhighd}. The only wormhole with a rigorously known field theory dual is the eternal AdS${}_2$ wormhole of \cite{wormads4}, shown in that paper to describe two coupled Sachdev-Ye-Kitaev (SYK) models. In higher dimensions, things are less clear, but for sure we have two strongly interacting field theories coupled via a double-trace coupling; an explicit example is given in \cite{wormfrei}.

Our goal is to examine what the nonlocal coupling and traversability from boundary to boundary means for dynamics, chaos and information transfer in dual field theory. Intimate relation of wormholes to quantum information and even the black hole entropy problem was found in \cite{erepr2}, in the framework of the celebrated ER=EPR proposal \cite{erepr1,worm2003} and the concepts of scrambling \cite{scramble,scramble2} and firewalls \cite{blacksmat}. The "diffusion of information" on the black hole horizon turns out to be related to the growth of time-disordered correlation functions known as out-of-time ordered correlators (OTOC) \cite{mss,butter,butterlocal,butterstring}. Essentially, OTOC diagnoses chaos during the equilibration of the quantum field theory/many body system after a perturbation.\footnote{This kind of chaos is essentially quasiclassical, and is only weakly related to the chaos in the deep quantum regime, described by the random matrix theory \cite{randmat}.} Another way of saying it is the scrambling concept of \cite{scramble}: the OTOC growth rate determines the timescale over which a small package of information distributes itself over macroscopic distances (over the whole black hole horizon in this case); supposing pure exponential growth $e^{\lambda t}$, the exponent $\lambda$ is usually dubbed the Lyapunov exponent, in analogy to classical Lyapunov exponents. Systems dual to a classical black hole saturate the MSS bound of \cite{mss}: $\lambda\leq 2\pi T$.

Where do the wormholes arise in the above story? The insight of \cite{erepr2} is that a particle passing from left to right infinity through a traversable AdS wormhole describes the teleportation from the left to the right subsystem of the dual field theory (actually, pair of field theories). The module of the commutator of two observables, one from the left and the other from the right subsystem, is clearly related to OTOC (for details see \cite{erepr2} and the discussion in the last section of this paper). One would therefore think that the OTOC behavior, i.e. the speed of the scrambling/the strength of chaos, will know about the opening of the wormhole. Coming to the GJW wormhole, we know that OTOC in the BTZ background at finite temperature exhibits maximum Lyapunov exponent $\lambda=2\pi T$; what happens when the double-trace coupling is turned on and a wormhole opens? Naively, we expect much slower chaos, i.e. much smaller $\lambda$: first, a wave packet on a black hole horizon interacts with the huge number of degrees of freedom inside and that is why it scrambles so quickly -- this is not the case for a wormhole, which has no horizon and its throat carries no internal degrees of freedom; second, a particle can go through the wormhole on the other side, and this means it does not stay forever in the interior, where the redshift is high and the scattering amplitude can grow very large.

Some work was done on OTOC calculations in various non-black hole systems. On the gravity side, \cite{otockevin} consider fuzzball and other microstate configurations alternative to black holes and find that indeed the scrambling slows down considerably in absence of a horizon. In \cite{otocpoojary} the authors find \emph{increased} chaos on the boundary of an AdS${}_3$-asymptotic geometry; \cite{otocfuzz} studies classical chaos in fuzzball backgrounds; anisotropic scrambling and space-dependent Lyapunov exponents are discussed in \cite{otocsarosi,otocsarosi2}. In wormhole backgrounds, to the best of our knowledge, mainly the field theory side was studied. Most relevant for us is the study \cite{otocsykexp} which considers two coupled SYK models (i.e. eternal AdS${}_2$ wormhole) and finds a Lyapunov exponent which is exponentially small in temperature $T$. Despite the mismatch in spacetime dimensions (our model being dual to a 1+1-dimensional theory), we will find something similar in one corner of the phase diagram, unfortunately in the regime where we haveleast control over the calculations. Further works on OTOC and phase transitions in SYK and related models include \cite{altmanfl,altmanphtr,aurelio,garcia,otocsach,otocsykgukitaev,otocsykzhang,sykrec}. In \cite{altmanfl} it is found that opening up a wormhole (coupling the two models) in general suppresses chaos.\footnote{A caveat is that the authors of that paper consider true quantum chaos, i.e. eigenvalue distribution, rather than the OTOC exponent.} The relation of dynamics and teleportation in wormholes are studied, e.g. in \cite{teleworm,teleswingle,telepapa,telebak}. Other recent works on wormholes and holography include \cite{jensen,deboer,antonini,nosaka,milekhin,godani,ahn,kiritsis,li,caceres,sarma,lunin,goto,iqbal}. In \cite{otocsarosi2,choi} other (non-wormhole) examples of non-black-hole metrics are considered, that generically give sub-maximal chaos. We will likewise find non-maximal exponents in the Lyapunov spectrum of wormhole geometries, although there is always also the fast, strongly chaotic mode, in addition to slower ones.

Operationally, we follow the tried recipe of OTOC calculation, explicated most saliently in \cite{butterstring}: we consider a bulk scalar field\footnote{One could, of course, consider some other field; we stick to the Klein-Gordon field for simplicity.} inserted at spatial infinity at time $t=0$ and another bulk scalar inserted at time $t>0$, scattering in the interior and reaching the other boundary. For a wormhole background however several complications arise. Not only is the geometry more complicated, but also time-dependent as the throat opens at some finite point in time; on top of that, the absence of infinite boost (present at the black hole horizon) invalidates some simple approximations which can be used for black holes. A very general method for calculating OTOC in coordinate representation, suitable also for time-dependent backgrounds, was found in \cite{balasubra2019otoc} (see also \cite{balasubra2019nohair} for a related work). However, for our wormhole this method turns out too complicated. We thus perform a perturbative calculation for a weak double-trace coupling, when the time-dependent nature of the geometry can be tackled perturbatively. Our approach is thus doubly perturbative: the double-trace coupling/wormhole throat size $\gamma$ is assumed to be small compared to the initial black hole mass $M$, and the infalling waves are assumed to have small spread (large momenta) so that the eikonal approximation holds. This is somewhat limiting of course, but we will show that it is good enough to gain some insight.

In Section \ref{sec2} we set the stage: first we quickly recapitulate the setup of opening a traversable wormhole via a double trace deformation, then we introduce some approximations that simplify the subsequent calculations and calculate the wormhole metric in the whole spacetime (in \cite{wormjaff} the metric is not given explicitly), and finally derive the bulk-to-boundary scalar propagators in this geometry. In Section \ref{sec3} we write the scattering amplitude for OTOC, discuss the complications arising from the time-dependent metric and absence of a horizon, and finally calculate the OTOC. Section \ref{sec4} describes the behavior of the Lyapunov exponents as a function of wormhole parameters; we also try to understand the result on field theory side. Section \ref{sec5} concludes the paper with a discussion of our findings in broader context. \ref{appa} and \ref{appb} bring some additional details of the calculations, and in \ref{appc} we consider the limit of a long-living (near-eternal) wormhole; this case is very interesting and brings some surprising results, but is much more difficult and we are forced to resort to very crude analytical estimates and the numerics; for this reason we give it separately from the main text.

\section{Setup: metric and Klein-Gordon equation in traversable AdS wormholes}\label{sec2}

\subsection{Simplified GJW wormholes and their metrics}\label{sec2worm}

Let us first briefly recapitulate from \cite{wormjaff} how the wormhole temporarily opens up and becomes traversable by a double-trace deformation. We start from the maximally extended BTZ black hole, containing two boundaries dual to two initially decoupled CFTs, thus describing a pure state through the thermofield dynamics (TFD) double $\sum_n e^{-\beta E_n}\vert n\rangle\vert n\rangle$, with $\beta=1/T$ the inverse temperature of the black hole and $\vert n\rangle$ being the CFT states with energy $E_n$ \cite{worm2003}, living in 1+1-dimensional spacetime with coordinates $(t,\phi)$. Now, \cite{wormjaff} couple the CFTs via the interaction term $\delta H$ in the Hamiltonian:
\bea
\nonumber &&H\equiv H_L-H_R\mapsto H+\delta H\\
\label{wormint}&&\delta H=-\gamma\int dt\int d\phi\chi(t,\phi)O_L(t,\phi)O_R(-t,\phi),
\eea
where $\gamma=\mathrm{const.}$ is the coupling strength, $\chi(t,\phi)$ encapsulates the spacetime dependence of the coupling and $O_{L,R}$ are some operators in the left and right CFT respectively. The first line defines the unperturbed Hamiltonian which contains two identical CFTs on the left- and right-hand side respectively. The holographic dictionary on double-trace deformations \cite{dbltrace1,dbltrace2} allows one to calculate the correction to the two-point correlation function. For a massive scalar of mass $m$ and conformal dimension $\Delta=1+\sqrt{1+m^2}$ (our case) this was done in \cite{wormjaff} for both the bulk-to-bulk ($G$) and bulk-to-boundary ($K$) propagator. The former allows one to express the wormhole-generating stress-energy tensor. To that end, we introduce the usual Kruskal coordinates:
\be
\label{kruskal}U/V=-e^{2r_ht},~(1-UV)/(1+UV)=r/r_h.
\ee
As usual, $r_h$ is the black hole horizon. Of course, the existence of the Kruskal coordinates hinges crucially on the black hole geometry, hence it is vital for our approach that the wormhole is not eternal so that we can always define the quantities in (\ref{kruskal}).\footnote{We will later consider a special, slow wormhole limit which has some properties of an eternal wormhole but it is still just a limit where the wormhole lifetime, while remaining finite, is larger than all other scales in the system.} Now the stress-energy tensor of the bulk scalar of conformal dimension $\Delta$ is obtained by definition as
\bea
\nonumber T_{UU}(U)&=&\lim_{U\to U'}\partial_U\partial_{U'}G(U,U')\\
\label{tmunuwh}G(t,t')&=&\gamma\sin\left(\pi\Delta\right)\int dt_1\chi(t_1)K(t'+t_1-\imath\beta/2)K(t-t_1)~+~(t\mapsto t').
\eea
Here, $G$ is the bulk propagator and $K$ is the zeroth-order (pure BTZ) bulk-to-boundary propagator. In the second line, $G$ is given as a function of time instead of $U$, but one can easily change the coordinates to Kruskal to insert $G(U,U')$ into the expression for $T_{UU}$. In comparison to (\ref{wormint}), we have put $\chi(t,\phi)\mapsto\chi(t)$ -- from now on we assume full isotropy in the angle $\phi$; this means we do not consider diffusion and spatial dependence of OTOC, but only time dependence, in a circular system (in other words, we consider spherical wormhole perturbations as in \cite{butter}). For the full derivation of the above result we refer the reader to \cite{wormjaff,wormbound}. The symmetry $T_{UU}(U)=T_{VV}(V)$ is exact at the horizon, where $V=0$ or $U=0$ holds. Away from the horizon (when both $U$ and $V$ are nonzero), $T_{UU}$ and $T_{VV}$ depend on both coordinates, however time reversal and parity symmetry lead to $T_{UU}(U,V)=T_{VV}(V,U)$. The resulting metric $g_{\mu\nu}$ now reads:
\be
\label{metricbtz}ds^2=-\frac{4L^2}{(1+UV)^2}dUdV+r_h^2\frac{(1-UV)^2}{(1+UV)^2}d\phi^2+\gamma h(U,V)dU^2+\gamma h(V,U)dV^2,
\ee
where $L$ is the AdS radius that we may put to unity and do so in the rest of the paper, and the symmetry of the stress-energy tensor implies the symmetry in the wormhole geometry, so that
$h_{UU}(U,V)=h_{VV}(V,U)\equiv h(U,V)\equiv\tilde{h}(V,U)$. The function $h$ is determined from the Einstein equations. Now from (\ref{tmunuwh}-\ref{metricbtz}) we can write down the one independent
Einstein equation and its solution:
\bea
\label{einseqwh}&&V\partial_Vh-U\partial_Uh-2h+2\gamma\frac{1-UV}{1+UV}T_{UU}(U)=0\\
\label{einseqwhsol}&&h(U,V)=\frac{2}{U^2}\frac{1-UV}{1+UV}\int_{U_0}^UdU'U'T_{UU}\left(U'\right).
\eea
In order to explicitly calculate $h$ from (\ref{einseqwhsol}), we need to insert a specific form of $\chi$ into (\ref{tmunuwh}). While \cite{wormjaff} brings the exact analytical solution for the average null energy, it is difficult to repeat their achievement for the metric itself; also, it would be convenient to have a solution (even if only approximate) given by an expression which is not too complicated (as the metric is only the starting point for later calculations); finally, the qualitative behavior of OTOC is likely not sensitive to the details of $T_{UU}$. For these reasons, we consider a rather drastic approximation, which was however already used in the literature and should have no
unphysical effects. We dub it the fast wormhole.

\emph{Fast wormhole.} We start from the idea given in \cite{wormbound}, where the double-trace coupling is made instantenous. In other words, we turn on the coupling at $t=t_0$ and turn it off at $t=t_f$, and then take the limit $t_f\to t_0$ ($U_f\to U_0$). This drastically simplifies the expressions for the average null energy, as found in \cite{wormbound}, and also for the metric as we will now see. We first need to find the stress-energy tensor in this approximation. To do that, we have to start from the defining expression (\ref{tmunuwh}) and plug in the expression for the bulk-to-boundary propagators in the BTZ background. But this is actually done in the original GJW calculation \cite{wormjaff} so we can just use their result. If the coupling is turned on at some $U_0$ and turned off at some $U_f$, the stress tensor is given by (Eq.~(3.9) in \cite{wormjaff}):
\bea
&&T_{UU}(U)=-\frac{4\gamma\Delta\Gamma\left(\frac{1}{2}\right)\Gamma(1-\Delta)}{\sqrt{2}\Gamma(3/2-\delta)}\times\nonumber\\
&\times&\lim_{U'\to U}\partial_U\int_{U_0}^{U_f} dU_1
\frac{F_1\left(1/2,1/2,\Delta+1,3/2-\Delta;\frac{U-U_1}{2U_1},\frac{U_1-U}{U_1(1+U'U_1)^2}\right)}{U_1^{-\Delta+1/2}(U-U_1)^{\Delta-1/2}(1+U_1U')^{\Delta+1}}.
\eea
The point $U'$ comes from point splitting in the calculation of the stress-energy tensor, and we have modified the limits of the integral (compared to \cite{wormjaff}) to study the interaction which is turned on for a finite time (from $U_0$ to $U_f$). Now we write $U_f=U_0+u$, expand in $u/U$ small and integrate. The outcome can be expressed in terms of the incomplete Euler beta function $B$:
\be
\label{quickt}T_{UU}(U)=2\gamma\Delta\sqrt{\pi}\frac{\Gamma(1/2-\Delta)}{\Gamma(1-\Delta)}B\left(-\frac{1}{U_0^2},1/2+\Delta,-2\Delta\right)(U-U_0)^{-2\Delta-1}\Theta(U-U_0),
\ee
For this stress tensor component $T_{UU}$, the Einstein equations at leading order in $\gamma$ (\ref{einseqwh}-\ref{einseqwhsol}) yield:
\bea
&&h(U,V)=2\gamma\Delta\sqrt{\pi}\frac{\Gamma(1-\Delta)}{\Gamma(3/2-\Delta)}\times\nonumber\\
&\times& B\left(-\frac{1}{U_0^2},1/2+\Delta,-2\Delta\right)(U-U_0)^{-2\Delta-1}\frac{1-UV}{1+UV}\Theta(U-U_0).\label{quickwhsol}
\eea
This is the same protocol as the one employed in \cite{wormbound}, except that we compute the metric itself and not the average null energy. Both quantities involve the integral of $T_{UU}$ but over different ranges: in (\ref{einseqwhsol}) we integrate from $U_0$ to a finite (arbitrary) $U$ whereas the average null energy is integrated for the whole geodesic, i.e. to $U\to\infty$. As a consequence, the integral and the limit $U_f\to U_0$ do not commute: if we started from the result for the average null energy in \cite{wormbound} and worked backwards to find $h$ consistent with it, we would find a different result.\footnote{For completeness we give it here. The $g_{UU}$ component of the metric, that we call $H$ for this case, reads $H(U,V)=8\Delta^2/(1-2\Delta)^2(U-U_0)^{-2\Delta-2}(1-UV)/(1+UV)\left(1+\left(2\Delta-1\right)\log U\right)$. Apart from the subleading logarithmic correction, this is
the same function form as (\ref{quickwhsol}) but with the power $-2\Delta-2$ instead of $-2\Delta-1$. There is nothing wrong with either result: as we have explained, taking the limit $U_f\to U_0$ in the geodesic average is not the same as taking it in the Einstein equation. One may speculate which limit would potentially be easier to realize in nature, but such questions are far from our current story.}

There is now a possible issue, hinted at also in \cite{wormbound}. The above expression for $h$ may diverge, if $T_{UU}(U\to 0)\sim U^\alpha$ with $\alpha\leq -2$. This is best seen by plugging (\ref{quickt}) into the Einstein equation (\ref{einseqwhsol}): it contains the integral $\int dU_1U_1T_{UU}(U_1)\sim \int dU_1(U_1-U_0)^{-2\Delta}$, which diverges at $U_1=U_0$ when $\Delta\geq 1/2$. A simple way to tackle this regime is the following. Observe first that for the marginal point $\Delta=1/2$, (\ref{quickwhsol}) is only logarithmically divergent and can be regularized. Regularizing as $\Delta\mapsto 1/2-\epsilon$ and expanding over $\epsilon$ to second order yields
\be
T_{UU}(U)\vert_{\Delta\to 1/2}=-\gamma\frac{\pi}{1+\Delta^2}\frac{1}{U_0}\frac{\epsilon}{\epsilon^2+(U-U_0)^2}\to-\tilde{\gamma}\frac{\delta(U-U_0)}{U_0}.
\ee
In other words, the stress-energy tensor itself (rather than the coupling $\chi$) becomes proportional to $\delta(U-U_0)$. In the second equality above we have introduced $\tilde{\gamma}\equiv\gamma\pi/(1+\Delta^2)$ and from now on we will write just $\gamma$ without the tilde as the constant factors can always be absosrbed in the definition of $\gamma$ which in our calculations will be a free parameter. This result, although special for $\Delta=1/2$, motivates us to \emph{assume} (there is no controlled way to formulate this, as the singularity in (\ref{quickwhsol}) is nonintegrable for $\Delta\geq 1/2$) that for $\Delta\geq 1/2$ the meaningful fast wormhole limit has the stress tensor
\be
\label{quicktdelta}T_{UU}(U)=-\gamma\frac{\delta(U-U_0)}{U_0}.
\ee
Once again, this is an \emph{assumption} -- for $\Delta>1/2$ there is no controlled way to define $T_{UU}(U)$ unless the point $U=U_0$ is excluded as this point is a nonintegrable singularity (see also a more detailed explanation in \cite{wormbound} and references therein). The difficulties come simply form the instantenous source model -- a realistic interaction would take a finite time. But we have found no inconsistencies stemming from this assumption and in fact the physical interpretation is obvious -- the pulse (instantenous) source gives rise to a shock wave of negative energy. The Dirac delta simply means we formally glue together two solutions.\footnote{A word of caution: here the wormhole-opening perturbation is itself a shock wave; this is distinct from the fact that the OTOC perturbation always contains a shock wave component, and in BH backgrounds, as we know \cite{butter}, OTOC is made solely from shock waves at leading order.} Adopting the form (\ref{quicktdelta}), the metric becomes
\be
\label{diracwhsol}h(U,V)=-\frac{2U_0}{U^2}\frac{1-UV}{1+UV}\Theta(U-U_0).
\ee
We encompass both the $\Delta\geq 1/2$ case and the $\Delta<1/2$ case under the name of fast wormhole.

It is also interesting to consider the opposite limit, when the wormhole is very long-living. This case can be called the \emph{slow wormhole} and it has some very interesting consequences for the main topic of the paper -- the behavior of OTOC. However, it also presents significant calculational difficulties which we could not fully resolve. For that reason, we have collected the (partial) results on the slow wormhole in \ref{appc}.

\subsubsection{Wormhole metric in radial coordinates}

For the solution of the Klein-Gordon equation and some other applications, it is convenient to have the wormhole metric also in $(t,r,\phi)$ coordinates. This is a harder nut to crack, and we could not obtain a closed-form expression. Instead, we match the solution in the throat region (small $r$) and the solution in the outer region (large $r$). The throat region develops near the black hole horizon and therefore should be close to an AdS metric (indeed, the asymptotically flat 3+1-dimensional wormhole studied in \cite{worm4} has AdS${}_2$ throat geometry). We introduce the new radial coordinate as
\be
\label{rho}\frac{r-r_h}{r_h}=\gamma\rho,
\ee
and consider the limit $\gamma\to 0,\rho\to\infty$ with $\gamma\rho\to\mathrm{const.}<1$; in other words, the mouth lives at large $\rho$ but the deviation of the throat metric from the BH metric is still finite and small because the wormhole opening scale $\gamma$ is assumed to be small. Since the metric is time-dependent, we need to consider different epochs in time separately. Remember that $g_{UU}=h$ and $g_{VV}=\tilde{h}$ are only significantly nonzero for certain times (or $U,V$ coordinates). We thus introduce the regimes (0) where $h,\tilde{h}$ are both negligible (I) only $h$ is significant (II) both $h$ and $\tilde{h}$ are significant and (III) only $\tilde{h}$ is significantly 
nonzero. Of course, the regime (0) is at leading order the same as the black hole metric. This picture becomes particularly simple for the Dirac delta model (\ref{diracwhsol}) where the regimes are sharply delineated by the step functions; the other cases can be found in Appendix A. In the regime (I) we have $U>U_0,V<U_0$, regime (II) is $U,V>U_0$ and the regime (III) has $U<U_0,V>U_0$. Actually, we can write $h$ explicitly as
\be
\label{diracwhh}h(t,r)=-2U_0e^{-2r_ht}r\frac{r+r_h}{r-r_h}.
\ee
Roughly speaking, (I) implies $t<t_0$, (II) means $t\approx t_0$ and in (III) we have $t>t_0$; these are really rough estimates as the Kruskal coordinates depend also on $r$. In each region we start from the solution (\ref{quickwhsol}) or (\ref{diracwhsol}) for $h$ (depending on $\Delta$), expanding in a series around $r=r_h$ for the throat region or around $r=\infty$ for the outer region.

Let us start from the throat region. The throat region is obtained by plugging in the solutions for $h,\tilde{h}$, passing to the Schwarzschild coordinates, transforming $r$ to $\rho$ as in (\ref{rho}) and then expanding around $\rho\to\infty$ and $\gamma\to 0$, i.e. expanding in a series in $1/\rho$ and $\gamma$. In the regime (I) we get
\be
\label{metricnearu}ds_I^2=\left(1+\gamma^2r_h^2\rho^2\right)\left(-dt^2+d\phi^2\right)+\frac{d\rho^2}{1+\gamma^2r_h^2\rho^2}+\frac{2\gamma U_0r_h^2}{1+\gamma^2r_h^2\rho^2}dtd\rho,
\ee
a nonstationary metric as it feels the onset of the perturbation. It has the form of a boosted AdS${}_3$. The boost comes, as we said, from turning on the left-right coupling, i.e. creating the wormhole dynamically, and AdS${}_3$ replaces the AdS${}_2$ throat of the eternal wormholes found in \cite{worm4} and \cite{wormfrei}; this is likely a consequence of our specific protocol of instantenous switching of the wormhole.

The regime (II) is stationary at leading order:
\be
\label{metricnearuv}ds_{II}^2=\left(1+\gamma^2r_h^2\rho^2\right)\left(-dt^2+d\phi^2\right)+\frac{d\rho^2}{1+\gamma^2r_h^2\rho^2}.
\ee
Notice that this AdS factor is \emph{not} merely a remainder of the original near-horizon AdS throat, as the latter is AdS${}_2\otimes\mathbb{R}$ for a BTZ black hole, and also our rescaled coordinate $\rho$ actually blows up for $\gamma=0$. As could be expected, the third region has the same form as (\ref{metricnearu}) with inverted time $t\to -t$. 

In the outer region, the solution must be close to the BTZ black hole. We consider again the same three regimes as before. Now we work directly with the $r$ coordinate, plugging in $h\left(U\left(t,r\right),V\left(t,r\right)\right)$ and expanding in $1/r$. In the regime (I) we get
\be
\label{metricfaru}ds_I^2=-(r^2-\tilde{r}_h^2)dt^2+\frac{dr^2}{r^2-\tilde{r}_h^2}+r^2d\phi^2+\frac{4\gamma r_h^2U_0}{r^2-r_h^2}dtdr,
\ee
where $\tilde{r}_h\equiv r_h(1-\gamma U_0)$. The regime (III) has identical metric except that $t\mapsto -t$. In the regime (II) the metric is
\be
\label{metricfaruv}ds_{II}^2=-(r^2-\tilde{r}_h^2)dt^2+\frac{dr^2}{r^2-\tilde{r}_h^2}+r^2d\phi^2.
\ee
In principle, the next step would be to match the metric solutions both "vertically", along the radial coordinate, for $\rho\to\infty$, and "horizontally", along the time axis. The outcome is a very cumbersome series expansion in $r-r_h$ or equivalently $1/\rho$. But fortunately we will not need it: what we want is the solution of the Klein-Gordon equation, and the matching between the solutions in different regions can be done directly for the Klein-Gordon scalar. We now proceed to the solution of the Klein-Gordon equation. We do that in each of the two regions (outer and inner) separately and then find the bulk-to-boundary propagator; in this way we need nothing beyond the asymptotic
metrics already found.

\subsection{Klein-Gordon equation in the wormhole background}\label{sec2kg}

The solution of the equation of motion for a scalar in wormhole background is an all-around useful goody to have for many purposes also outside the scope of this paper (stability analysis, equilibrium correlation functions and spectral functions, etc). We first solve the Klein-Gordon equation in each region (outer, throat) and in each regime (I,II,III) separately. The "horizontal" matching over time can then easily be done directly, and matching along $r$ requires a series expansion. The case $\Delta\geq 1/2$ is again the simplest so let us again show this case in the main text; the other cases we describe in \ref{appa}. For all solutions we choose the boundary conditions appropriate for the bulk-to-boundary propagator: it should be well-behaving in IR (whereas in the UV it reflects the presence of a Dirac delta source, i.e. has a non-normalizable mode). Therefore, the solution has to be finite for
$\rho\to 0$ or equivalently for $r\to r_h$. Finally, in all equations we disregard quadratic and higher order terms in $\gamma$, as we do in the whole paper.

\subsubsection{Inner region}

In the throat region, the AdS-like geometry leads to Bessel functions in the solutions, as could be expected. Expanding over the energy eigenvalues $\omega$ and angular momentum eigenvalues $\ell$, the equation of motion for the wavefunction $\Phi(t,\rho,\phi;\omega,\ell)=\exp(-\imath\omega t+\imath\ell\phi)\tilde{\Phi}(\rho)$ reads:
\be
\label{kgthroateq}\tilde{\Phi}''(\rho)+\left(\frac{3}{\rho}-\frac{4\sigma\imath\gamma U_0\omega r_h^2}{\rho^4}\right)\tilde{\Phi}'(\rho)+\frac{\omega^2-\ell^2-m^2\rho^2}{\rho^4}\tilde{\Phi}(\rho)=0,
\ee
where $\sigma=-1,0,1$ for regimes (I,II,III), respectively. In the regime (II) we easily get
\bea
\nonumber\Phi^{\mathrm{throat}}_{II}(t,\rho,\phi;\omega,\ell)&=&\frac{1}{\rho}e^{-\imath\omega t+\imath\ell\phi}K_{\sqrt{1+m^2}}\left(\sqrt{\ell^2-\omega^2}\rho\right),~~\omega<\ell,\\
\label{kgthroatii}\Phi^{\mathrm{throat}}_{II}(t,\rho,\phi;\omega,\ell)&=&\frac{1}{\rho}e^{-\imath\omega t+\imath\ell\phi}J_{\sqrt{1+m^2}}\left(\sqrt{\omega^2-\ell^2}\rho\right),~~\omega>\ell.
\eea
In the other two regimes we elliminate the extra term in (\ref{kgthroateq}), proportional to $\sigma=\pm 1$, by transforming:
\be
\label{kgthroati}\Phi^{\mathrm{throat}}_{I,III}(t,\rho,\phi;\omega_{I,III},\ell)=c_{I,III}\Phi^{\mathrm{throat}}_{II}(t,\rho,\phi;\omega_{I,III},\ell)\exp\left(\frac{\mp 2\imath\gamma U_0\omega_{I,III} r_h^2}{3\rho^3}\right).
\ee
By explicitly denoting the frequency in the regime I/III by $\omega_{I,III}$, we have emphasized the fact that any value for frequency is allowed \emph{a priori} for each regime in isolation (we denote $\omega_{II}\equiv\omega$). For the solution in the whole range, we in general match the modes with different frequencies in different regimes, because the background is time-dependent and a harmonic function $e^{-\imath\omega t}$ is not even a leading order approximate solution for all times.  The
formulas (\ref{kgthroatii}-\ref{kgthroati}) complete the solution in each regime separately. We first need to perform the "horizontal" matching, along the time axis, and afterwards the "vertical" matching, along $r$. The horizontal matching is done at $t=t_0$ (because the points $\pm t_0$ separate the three regimes) and at fixed $\rho$. Let us do this in detail for the region II/region III match. Equating (\ref{kgthroatii}) and (\ref{kgthroati}) at the mouth, i.e. $\rho=1$ (or $r=r_h(1+\gamma)$ in the original coordinates) gives the condition (notice that the $\rho$-dependent factors cancel out):
\be
\label{kgmatchiii}\exp\left(-\imath\omega t\right)=c_{III}\exp\left(-\imath\omega_{III}t+\frac{2\imath\gamma U_0\omega_{III} r_h^2}{3}\right).
\ee
The second condition comes from expanding (\ref{kgthroatii}) and (\ref{kgthroati}) in $\rho$ large and equating the results:
\be
\label{kgmatchiii2}\rho^{-1+\sqrt{1+m^2}}\left(\ell^2-\omega^2\right)^{-\frac{1}{2}\sqrt{1+m^2}}=
c_{III}\rho^{-1+\sqrt{1+m^2}}\left(\ell^2-\omega_{III}^2\right)^{-\frac{1}{2}\sqrt{1+m^2}}.
\ee
The system (\ref{kgmatchiii}-\ref{kgmatchiii2}) yields the solutions:
\be
c_{III}=1+\delta\omega\frac{\omega\sqrt{1+m^2}}{\ell^2-\omega^2},~\omega_{III}=\omega+\delta\omega,~~\delta\omega=2\gamma U_0
\ee
The regime I is analogous, with the sign of $\delta\omega$ inverted, so the matched solution in the inner region for all times reads:\footnote{Remember we disregard higher-order corrections in $\gamma$.}
\bea
\nonumber&&\Phi^{\mathrm{throat}}(t,\rho,\phi;\omega,\ell)=
\left(1+2\gamma U_0\sqrt{1+m^2}\frac{\omega\mathrm{sgn}(t-t_0)}{\ell^2-\omega^2}\right)\times\\
&&\times\frac{1}{\rho}e^{-\imath\omega t+\imath\ell\phi-\frac{2\imath\gamma U_0\omega r_h^2}{3\rho^3}\mathrm{sgn}(t-t_0)}K_{\sqrt{1+m^2}}\left(\sqrt{\ell^2-\omega^2}\rho\right).\label{kgthroat}
\eea
The matching is done for $\omega<\ell$; for $\omega>\ell$ everything is of course the same just with $K\mapsto J$. The matched solution is of the same form as (\ref{kgthroati}), and for small $\omega$ it yields precisely (\ref{kgthroatii}) at leading order, as it should be.

\subsubsection{Outer region}

Now consider the outer region. The Klein-Gordon equation reads
\bea
&&\phi''(r)+\frac{3r^2-\tilde{r}_h^2-\frac{4\sigma\imath\gamma U_0\omega\tilde{r}_h^2}{r}}{r(r^2-\tilde{r}_h^2)}\phi'(r)+\nonumber\\
&&+\frac{r^2\left(\omega^2-m^2\left(r^2-\tilde{r}_h^2\right)-2\sigma\imath\gamma U_0\omega\tilde{r}_h^2\frac{r^2-\tilde{r}_h^2}{r}\right)-\ell^2(r^2-\tilde{r}_h^2)}{r^2(r^2-\tilde{r}_h^2)}\phi(r)=0,\label{kgouteq}
\eea
where $\sigma$ has the same meaning as before ($-1,0,1$ for the regimes I,II,III), and $\tilde{r}_h\equiv r_h(1-\gamma U_0)$ as in (\ref{metricfaru}). In the regime (II) the solution is the hypergeometric function, as already found, e.g. in \cite{propbtz1,propbtz2}:
\bea
\nonumber&&\Phi^{\mathrm{out}}_{II}(t,r,\phi;\omega,\ell)=e^{-\imath\omega t+\imath\ell\phi}r^{\imath\tilde{\omega}-\Delta}(r^2-\tilde{r}_h^2)^{-\imath\tilde{\omega}/2}
{}_2F_1\left(a,b,\Delta;\frac{r^2}{\tilde{r}_h^2}\right)\\
\label{kgoutii}&&a=\frac{\tilde{\ell}}{2}-\frac{\tilde{\omega}}{2}+\frac{\Delta}{2},~b=-\frac{\tilde{\ell}}{2}-\frac{\tilde{\omega}}{2}+\frac{\Delta}{2}.
\eea
We have introduced the notation $\tilde{\ell}\equiv\ell/\tilde{r}_h$, $\tilde{\omega}\equiv\omega/\tilde{r}_h$, and picked the branch that remains smooth in the interior (for $r\to \tilde{r}_h$), which is appropriate for the Feynmann propagator. In the regimes (I,III) we can again reduce the equation to the $\sigma=0$ case by introducing
\bea
&&\Phi^{\mathrm{out}}_{I,III}(t,r,\phi;\omega_{I,III},\ell)=
C_{I,III}\Phi^{\mathrm{out}}_{II}(t,r,\phi;\omega_{I,III},\ell)\times\nonumber\\
&&\times\exp\left[\mp\imath\gamma U_0\left(\frac{\omega r-\tilde{r}_h^2}{r^2-r_h^2}-\omega\arctan\frac{r}{r_h}\right)\right].\label{kgouti}
\eea
For the matching we equate (\ref{kgoutii}) to (\ref{kgouti}) at $r\to\infty$, resulting in:
\bea
\nonumber C_{III}&=&\exp\left(\frac{\imath\gamma U_0\left(2\imath\omega\pi+\pi r_h\gamma U_0-2(\omega-\imath r_h)U_0\right)}{2(2-\gamma U_0)r_h}\right),~\omega_{III}=\omega+\delta\omega\\
\delta\omega&=&\frac{\gamma U_0(\omega-\imath r_h)}{2-\gamma U_0}.
\eea
Plugging back into (\ref{kgoutii}-\ref{kgouti}) gives the matched solution:
\bea
\nonumber&&\Phi^{\mathrm{out}}(t,r,\phi;\omega,\ell)=\left(1+2\imath\gamma U_0\frac{\omega\mathrm{sgn}(t-t_0)}{\sqrt{\omega^2-\tilde{r}_h^2/r^2}}\right)\exp\left[\imath\gamma\omega U_0\mathrm{sgn}(t-t_0)\left(\frac{r}{r^2-r_h^2}-\arctan\frac{r}{r_h}\right)\right]\times\\
\label{kgout}&&\times
e^{-\imath\omega t+\imath\ell\phi}r^{\imath\tilde{\omega}-\Delta}(r^2-\tilde{r}_h^2)^{-\imath\tilde{\omega}/2}{}_2F_1\left(\frac{\imath\tilde{\ell}-\imath\tilde{\omega}+\Delta}{2},\frac{-\imath\tilde{\ell}-\imath\tilde{\omega}+\Delta}{2},\Delta;\frac{r^2}{\tilde{r}_h^2}\right),
\eea
where we again disregard higher-order terms in $\gamma$. This concludes the solution of the Klein-Gordon equation. Now we will feed these results into the bulk-to-boundary propagator.

\subsubsection{Bulk-to-boundary propagator}
 
To remind, the bulk-to-boundary propagator $K(r;\mathbf{x},\mathbf{x}')$ satisfies the homogeneous equation of motion in the bulk and behaves as the Dirac delta at the boundary when properly rescaled:
$\lim_{r\to\infty}r^{D-\Delta}K(r;\mathbf{x},\mathbf{x}')=\delta(\mathbf{x}-\mathbf{x}')$. Analytical expressions both for bulk-to-bulk and bulk-to-boundary propagators in the BTZ black hole background are known \cite{propbtz2}. For the latter, it reads
\be
\label{kbnd}K_\mathrm{BTZ}(r;t,0;\phi,0)=\frac{r_h^{2\Delta}}{2^{\Delta+1}\pi}\left(r\cosh r_h\phi-(r^2-r_h^2)^{1/2}\cosh r_ht\right)^{-\Delta}.
\ee
Translation invariance allows putting $t'$ and $\phi'$ to zero. The result for $K_\mathrm{BTZ}$ is usually obtained from the bulk-to-bulk propagator $G(r,r';\mathbf{x},\mathbf{x}')$, which is in turn obtained through the method of images from pure AdS${}_3$ spacetime. Since the
wormhole is not simply related to pure AdS${}_3$ anymore, it is awkward to use the method of images. We instead construct $G$ by definition, from the eigenmodes, and then find $K$ at leading order by expanding around $r'\to\infty$. The defining expression for $G$ is
\be
\label{gsum}G(r,r';t,t';\phi,\phi')=\sum_\ell\int d\omega\Phi^{\mathrm{out}}(t',r',\phi';\omega,\ell)\Phi^{\mathrm{throat}}(t,\rho(r),\phi;\omega,\ell).
\ee
The modes $\Phi^{\mathrm{throat}}$ and $\Phi^{\mathrm{out}}$ satisfy the physical boundary conditions in the interior (smooth solution) and at the boundary (converging to the pure AdS solution), respectively. Therefore, conveniently, for the mode sum we need only the throat and outer region solutions (\ref{kgthroat},\ref{kgout}), not the matched solution in the whole spacetime. Notice that $\omega$ is just a parameter, not the frequency in the true sense, as the equations are time-dependent and the solutions, as we have found, are not harmonic in time. The actual steps needed to perform the sum are given in \ref{appb}. When this is done, we exploit the connection between $G$ and the bulk-to-boundary propagator: $K(r;t,t';\phi,\phi')=\lim_{r'\to\infty}r'^{\Delta}G(r,r';t,t';\phi,\phi')$ (see e.g. \cite{wormjaff}). For further use, it will be most convenient to use the mixed-coordinate-system propagator where $r'$ and $t'$ remain but $r$ and $t$ are transformed to $U$ and $V$. This leads us to the following result at leading order in $\gamma$:
\bea
\nonumber K(U,V;t';\phi,\phi')&=&
\left[1-\frac{2\gamma UV}{1+UV}\frac{1}{\gamma^2+\left(\frac{2UV}{1+UV}\right)^2\left[e^{-\tilde{r}_ht'}U-e^{\tilde{r}_ht'}V+\cosh\left(\tilde{r}_h\left(\phi-\phi'\right)\right)\right]^2}\right]^\Delta\times\\
\label{kbnddirac}&\times &e^{\frac{\gamma U_0}{4}}\left[e^{-\tilde{r}_ht'}U-e^{\tilde{r}_ht'}V+\cosh\left(\tilde{r}_h\left(\phi-\phi'\right)\right)\right]^{-\Delta}.
\eea
This is the final step of this rather cumbersome calculation. The case $\Delta<1/2$ is even worse; the algebra is very tedious. We have performed all the series expansions in the Mathematica package and have not bothered to simplify all the intermediate expressions into a humanly readable form.\footnote{We are ready to provide the Mathematica notebooks to interested readers.} We give the outcome in \ref{appa}. In the main text we will just make use of them to give the leading order results for the scattering amplitude, i.e. OTOC itself, where the expressions are a bit simpler. Finally, since we ignore the
diffusion in $\phi$, we could put $\phi=\phi'=0$ from the beginning but we prefer to have the most general form of the propagator for possible later use.

\section{The scattering amplitude and OTOC}\label{sec3}

\subsection{The definition of OTOC}\label{sec3def}

Let us first remind ourselves of the definition of OTOC and its connection to the bulk scattering amplitude. We may motivate the out-of-time ordered correlation function by noticing that the module of the commutator of some operator $O_1$ at time $t_1=t$ and $O_2$ at time $t_2=0$ contains both time-ordered and time-disordered quantities:
$\langle\vert [O_1(t),O_2(0)]\vert^2\rangle=2\langle(O_1^\dagger(t)O_2^\dagger(0)O_1(t)O_2(0)\rangle+2\langle O_1^\dagger(t)O_1(t)O_2(0)O_2^\dagger(0)\rangle$, taking into account the invariance of the expectation value to cyclic permutations. While the second term, the time-ordered correlator (TOC) presumably factorizes at long times, the first, the OTOC term, does not.\footnote{Some authors use the term OTOC for the expectation value of the whole commutator $\langle\vert [O_1(t),O_2(0)]\vert^2\rangle$. We however reserve the term OTOC for the second term, whereas the first term, which is time-ordered, is called TOC.} The two-time commutator itself can be interpreted as the perturbation of the operator $O_2$ upon evolving the system forward for time $t$, acting on it by the operator $O_1$, and then evolving backwards for time $-t$, in analogy to the Loschmidt echo. For a more detailed physical discussion we refer the reader, e.g. to \cite{butterstring,otocsach}. In this paper we
focus on the combination $O_1^\dagger(t)O_2^\dagger(0)O_1(t)O_2(0)$ and call it OTOC.

But for a wormhole, the above definition of OTOC becomes subtle. For the familiar calculation in the black hole background \cite{butterstring}, one works in the maximally extended black hole spacetime in Kruskal coordinates which has two boundaries and two CFTs. Therefore, a field theory operator $O_1$ can be from the left or right CFT, $O_{1L}$ or $O_{1R}$, so the definition of OTOC has to specify which operators we consider and various combinations are possible, like $\langle O_{1L}^\dagger(t)O_{2R}^\dagger(0)O_{1L}(t)O_{2R}(0)\rangle$ or $\langle O_{1L}^\dagger(t)O_{2L}^\dagger(0)O_{1L}(t)O_{2L}(0)\rangle$ etc. However, the left and right operators are related in an easy way in the black hole
background, so all possible left/right combinations for the OTOC are related by analytic continuation, by adding $\pm\imath\beta/2$ to the time argument of the right-hand operators. This ceases to be true for a wormhole, where there are genuinely two different, entangled CFTs. In this case, the two-sided OTOC includes some (\emph{a prirori} unknown) operation that translates, e.g. $O_{1R}^\dagger(t)$ to $O_{1L}^\dagger(t)$. The only function that we know how to calculate without introducing any new hypotheses about the structure of the boundary action is the fully one-sided OTOC. This is the object we calculate in this paper:
\be
\langle O_{1R}^\dagger(t)O_{2R}^\dagger(0)O_{1R}(t)O_{2R}(0)\rangle.
\ee
From now on we will drop the index $R$ as it is understood for every operator we consider. This is also the function which is most naturally related to the field-theory and many-body applications, including the pioneering work by Larkin and Ovchinnikov where such objects were first studied.

\subsection{Setting the stage}\label{sec3set}

Now we set to calculate the quantity $\langle O_1^\dagger(t)O_2^\dagger(0)O_1(t)O_2(0\rangle$ which from now on we will also denote by $D(t,0)$ for the sake of brevity. From now on we take $O_1$ and $O_2$ to be operators dual to the bulk scalar (Klein-Gordon) fields with different conformal dimensions $\Delta_1$ and $\Delta_2$. The wormhole-opening perturbation is generated by the field with dimension $\Delta_1$, i.e. the conformal dimension $\Delta$ from the previous section is $\Delta_1$ in the OTOC setup.\footnote{In principle, the wormhole-generating perturbation could be due to an altogether different field, with dimension distinct from both $\Delta_1$ and $\Delta_2$, but apparently we do not lose in generality by taking it equal to $\Delta_1$.} As explicated in \cite{butter,butterlocal,butterstring}, the fundamental holographic relation connecting the OTOC to a bulk scattering amplitude can be schematically represented as
\bea
\nonumber &&D(t,0)=\int dP\Psi_1^\dagger(t,\phi_1;P)\Psi_2^\dagger(0,\phi_2;P)\Psi_1(t,\phi_1;P)\Psi_2(0,\phi_2;P)\langle\mathrm{IN}(P)\vert\mathrm{OUT}(P)\rangle\nonumber\\
\label{otoc}&&\vert\mathrm{OUT}(P)\rangle=\vert\mathrm{IN}(P)\rangle e^{\imath S_c(P)},
\eea
where $P$ is any set of variables that characterizes the IN and OUT states (for a BH these are the conserved momenta $p$ and $q$ of the fields $\Psi_1$ and $\Psi_2$ respectively), and the eikonal approximation implies that the phase shift equals the classical action $S_c$. The bulk wavefunctions $\Psi_{1,2}$ with masses $m_{1,2}$ are dual to the field theory operators $O_{1,2}$ with conformal dimensions $\Delta_{1,2}$.

Our strategy will be to treat the wormhole opening $\gamma$ perturbatively, so the whole OTOC calculation that we perform essentially builds up on the calculation in the BH background. Let us thus first remind the reader how it works for a BH, emphasizing those points which are going to change when the wormhole opens. The wavefunctions $\Psi_{1,2}$ can be represented as Fourier transforms of the coordinate wavefunctions obtained with the help of the bulk-to-boundary propagators for the bulk fields dual to operators $O_1,O_2$, so we can write
\bea
\Psi_1(t,\phi_1;p)&=&\int dU\int dV\int d\phi'e^{\imath\frac{g_{UV}}{2}\left(p^VU+p^UV\right)}K_1^{\mathrm{(BH)}}(U,V;t;\phi_1,\phi')\delta(U)\label{otocfuns1}\\
\Psi_2(0,\phi_2;q)&=&\int dU\int dV\int d\phi'e^{\imath\frac{g_{UV}}{2}\left(q^VU+q^UV\right)}K_2^{\mathrm{(BH)}}(U,V;0;\phi_2,\phi')\delta(V),~~~\label{otocfuns2}
\eea
where $p,q$ are the momenta of the fields $\Psi_1$ and $\Psi_2$, $p^V,p^U$ are the components of $p$ conjugate to the coordinates $U,V$ (and likewise for $q^V,q^U$), and we have emphasized that the bulk-to-boundary propagators are for the BH background, not our WH propagators (\ref{kbnddirac}). The sources (bulk initial configurations) on which the propagators act are the geodesics of the infalling and outgoing trajectory, defined by $U=0$ and $V=0$ respectively, hence the Dirac deltas in (\ref{otocfuns1}-\ref{otocfuns2}). In the literature one usually immediately puts $U=0$ or $V=0$ when writing (\ref{otocfuns1}-\ref{otocfuns2}) but we deliberately want to write it in a way which paves the road to the generalization for the wormhole. Also, since we only consider spherical perturbations in this paper, the angular dependence and the angular integrals drop out and we will not write them from now on. Inserting (\ref{otocfuns1}-\ref{otocfuns2}) into the amplitude (\ref{otoc}), we get
\be
D(t,0)=\int dp^Up^U\int dq^Vq^Ve^{\imath S_c\left(p^U,q^V\right)}\Psi_1^\dagger(t;p^U)\Psi_2^\dagger(0;q^V)\Psi_1(t;p^U)\Psi_2(0;q^V).\label{otocint}
\ee
Notice that the four wavefunctions $\Psi_1^\dagger,\Psi_2^\dagger,\Psi_1,\Psi_2$ have four double integrals over the coordinates $U$ and $V$, since each of the wavefunctions is Fourier-transformed from the coordinate to the momentum representation; so we need to perform four integrals $\int dU\int dV$ (which is easy for a BH as the Dirac deltas immediately kill half of the integrals) and then the momentum integral $\int dp^U\int dq^V$. In order to complete the calculation, we need to supply the classical, i.e.
on-shell action, which is obtained in \cite{butterstring} as
\be
S_c=\frac{1}{4}\int d^3x\sqrt{-g}\Delta g_{\mu\nu}t^{\mu\nu},\label{sclass}
\ee
where $t^{\mu\nu}$ is the stress-energy tensor for a point particle (this is the eikonal approximation), and $\Delta g_{\mu\nu}$ is the shock-wave perturbation of the metric caused by the propagating OTOC field (computed also in the eikonal approximation). In a non-BH geometry, the metric perturbation will in general contain not only the shock wave $\Delta g_{\mu\nu}$ but also a non-shock-wave contribution.

Let us now sit back and think what will change when we try to follow the same path for a wormhole. The general formula (\ref{otoc}) remains. Different geometry however means different solutions to the Klein-Gordon equations, different bulk-to-boundary propagators $K$ in (\ref{otocfuns1}-\ref{otocint}), different geodesics giving rise to different stress-energy tensor $t^{\mu\nu}$ in (\ref{sclass}) and thus also different backreaction on the metric in the same equation for $S_c$. In detail, this means the following:
\begin{enumerate}
\item The geodesic $X_\mu(\tau)$ that enters the stress-energy tensor $t_{\mu\nu}$ will differ from the BH geodesic mainly for small $U$ or small $V$ -- this is where falling into the horizon is replaced by the tunnelling through the wormhole throat. Locally, this is a perturbative effect linear in $\gamma$. Global effects, due to different global shape of the trajectory, are considered in \ref{appc}, as they are most relevant for the long-living slow wormholes. Here we take into account only the local perturbative effect.
\item The WH geometry is explicitly time-dependent, thus the momenta are not conserved anymore, and $p,q$ are not well-defined quantum numbers in (\ref{otoc}-\ref{otocint}). However, the momentum nonconservation can also be treated perturbatively in $\gamma$ so we can write the momenta as $p=p_\mathrm{initial}+\gamma(\ldots)$ -- the sum of the asymptotic momentum and the wormhole-induced correction. This will influence the stress-energy tensor of the perturbation $t_{\mu\nu}$ as well as the amplitude
calculation from Eq.~(\ref{otocint}) and can have drastic consequences: the extra terms in $S_c$ stemming from the change in momentum $\gamma(\ldots)$ can make the classical action nonquadratic in the center-of-mass momentum, thus leading to non-maximal chaos as opposed to the BH case \cite{butterstring}.
\item The backreaction will be more complicated than just a shock wave. The perturbation of the metric will be of the form shock wave $\Delta g_{\mu\nu}$ (in the eikonal approximation there is always a shock wave contribution because point particles and rings always source a shock-wave metric \cite{thooft1,thooft2,sfetsos}) plus a smooth correction $\delta g_{\mu\nu}$. 
\item Different propagators $K$ will change the values of the scattering amplitude but it will turn out they do not lead to any qualitative changes from the BH case.
\end{enumerate}
The exciting things happen as a consequence of (1) and (2) in the above list, and now we describe how this happens. We first write the geodesic equations and solve them for $UV$ small, and then we consider the large-scale geometry of geodesics; these results allow us to write the stress-energy tensor in the eikonal approximation. The second step will be the calculation of the backreaction, resulting in the metric correction $\Delta g_{\mu\nu}$. Then it is easy to compute the on-shell action, and the final step is the calculation of the scattering amplitude from the ingredients previously obtained.

\subsection{The calculation}\label{sec3calc}

\subsubsection{The geodesic equation}

\emph{Near-mouth behavior.} In BH background, $U=0$ is a geodesic, and in that case $q^V\equiv q$ appearing in (\ref{otocfuns1}) is the only nonzero component of the momentum (and likewise $p^U\equiv p$ in (\ref{otocfuns2})). Clearly, this is not the case for a wormhole. Therefore, $q^U$ and $p^V$ are both nonzero, and the metric recieves corrections in $UU$, $VV$, $UV$ and $\phi\phi$ components both from incoming and outgoing waves (the remaining components are zero by symmetry). To see that, start from the equation for the radial geodesic:
\bea
&&V''+\gamma hg^{UV}U''+\gamma g^{UV}\partial_VhU'V'+\nonumber\\
\label{geogen}&&+g^{UV}\partial_Vg_{UV}\left(V'\right)^2+\frac{1}{2}\gamma g^{UV}\left(\partial_Uh\left(U'\right)^2-\partial_U\tilde{h}\left(V'\right)^2\right)=0,
\eea
where $U=U(\tau)$, $V=V(\tau)$ and also $h=h(U(\tau),V(\tau))$ and $\tilde{h}=h(V(\tau),U(\tau))$; by $g_{UV}$ we denote the component of the unperturbed wormhole metric (\ref{metricbtz}). The second equation is equivalent to the above, with $U\leftrightarrow V$.\footnote{This also means $h\leftrightarrow\tilde{h}$.} For $U\ll 1$ (and for $V$ small analogously), we can expand the equation (\ref{geogen}) about the BH geodesic $U=0,V=\tau$\footnote{Of course, the form of $V(\tau)$ depends on the gauge choice but we can always pick the gauge where $V=\tau$.} quadratically in $\gamma$. After some algebra, the solution reads:
\bea
\label{geodeq1}U(\tau)&=&\frac{\gamma}{4}\int_0^\tau d\tau'\int_0^{\tau'}d\tau''\partial_Uh\left(V\left(\tau''\right),U\left(\tau''\right)\right)+O\left(\gamma^3\right)\\
\label{geodeq2}V(\tau)&=&\int_0^\tau\frac{d\tau'}{1+\frac{\gamma}{4}\int_0^{\tau'}d\tau''\partial_Vh\left(V\left(\tau''\right),U\left(\tau''\right)\right)}+O\left(\gamma^3\right)=\tau+O\left(\gamma\right).
\eea
In the second line we have emphasized that we only need the first-order $\gamma$ correction of the BH geoedesic, as the subleading corrections would only influence third- and higher-order corrections to the scattering amplitude. Inserting the relation (\ref{einseqwhsol}) between $h$ and the stress tensor, we further get
\bea
\label{geodeqh1}U(\tau)&=&\frac{\gamma}{2}T_{UU}(0)\tau-\gamma\int_0^\tau d\tau'\tau'^2h(\tau',0)+O\left(\gamma^3\right)\\
\label{geodeqh2}V(\tau)&=&\tau+\frac{\gamma}{12}\left(T_{UU}(0)\tau^3-T'_{UU}(0)\tau^2\right)+O\left(\gamma^3\right).
\eea
Now inserting a specific wormhole model, in our case (\ref{quickwhsol}) or (\ref{diracwhsol}), we obtain the equation for the geodesic (the trajectory equation) with the first-order correction in $\gamma$:
\be
\label{eqorbit}U-\gamma f(V)=0,~~f(V)=\int_0^V d\tau'\tau'^2 h(\tau',0)+O\left(\gamma^3\right).
\ee
The $V$-dependence is thus implicit, solely though the upper limit of the integral, but for our model cases (i.e. for our solutions for $h$), it was easy to write down explicitly.
The expansion (\ref{geodeqh1}-\ref{geodeqh2}), valid in the region of $UV$ small, precisely where the redshift is the highest (of order $O(1/\gamma)$), captures the leading \emph{local} backreaction effect, and we will shortly calculate the leading contribution to the stress-energy tensor. The \emph{global} difference from the BH case -- the fact that the orbit continues through the throat -- does not matter at leading order as the backreaction deep in the throat and further is subleading. 

\emph{Stress-energy tensor.} Now we can calculate the stress-energy tensor of the infalling wave by definition. The total backreaction is due to two incoming and two outgoing waves; we can write the equations for one incoming wave and in the end add the contributions from the other waves obtained by symmetry $U\leftrightarrow V$. Starting from the single-particle action $S_p=-\int d\tau\sqrt{-g_{\mu\nu}\dot{X}^\mu\dot{X}^\nu}$, with $X_\mu=(U,V,)$, we get:
\be
\label{tmunucp}t_{\mu\nu}=\frac{2}{\sqrt{-g}}\frac{\delta S_p}{\delta g_{\mu\nu}}=\frac{1}{\sqrt{-g}}\delta\left(U-U\left(\tau\right)\right)\frac{g_{\mu\alpha}g_{\nu\beta}\dot{X}^\alpha\dot{X}^\beta}{\dot{U}(\tau)}
=\left(\begin{matrix} t_{UU} & t_{UV} & 0 \\ t_{UV} & t_{VV} & 0 \\ 0 & 0 & 0 \end{matrix}\right),
\ee
The components $t_{UV},t_{VV}$ are due to the wormhole opening and are proportional to the wormhole size $\gamma$. The (non-conserved) components $p^U(\tau),p^V(\tau)$ of the momentum $p$ read
\be
\label{geodp}p^V(\tau)=-\gamma p^U(\tau=0)g_{UV}h\left(U\left(t\right),V\left(t\right)\right)+O\left(\gamma^3\right),~~p^U(\tau)=p^U(\tau=0)+O\left(\gamma^3\right),
\ee
so we are in fact somewhat lucky: even to second order the $p^U$ momentum is still approximately constant. From now on we denote the initial (asymptotic) momentum by $p_0\equiv p^U(\tau=0)$, and $q_0\equiv q^V(\tau=0)$ for the other wave. From (\ref{tmunucp}) and (\ref{geodp}) we get
\bea
\label{tmunucuu}t_{UU}(U,V)&=&p_0\delta\left(U-\gamma f\left(V\right)\right)\frac{g_{UV}^2}{\sqrt{-g}}\left(1+\gamma^2\left(g^{UV}\right)^2h\tilde{h})\right)\\
\label{tmunucvv}t_{VV}(U,V)&=&p_0\delta\left(U-\gamma f\left(V\right)\right)\frac{\gamma^2\tilde{h}^2}{4\sqrt{-g}}\\
\label{tmunucuv}t_{UV}(U,V)&=&p_0\delta\left(U-\gamma f\left(V\right)\right)\frac{\gamma g_{UV}\tilde{h}}{2\sqrt{-g}}\left(1-\frac{\gamma^2\left(g^{UV}\right)^2h\tilde{h}}{4}\right),
\eea
where the argument of the Dirac delta is the new trajectory equation. In line with our perturbative treatment, we expand (with help of (\ref{geodeqh1})):
\be
\delta\left(U-\gamma f\left(V\right)\right)=\delta(U)-\gamma\int_0^\tau d\tau'\tau'^2h(\tau',0)+O(\gamma^3),
\ee
which suffices to obtain the backreaction to second order. The contribution of the other wave is obtained, as we said, by exchanging $U$ and $V$.

\subsubsection{Backreaction: shock wave and beyond}

The next step is the backreaction of the stress tensor (\ref{tmunucuu}-\ref{tmunucuv}). As we already mentioned, the presence of the Dirac delta in the stress-energy tensor, inherent to the eikonal approximation, means that the metric change $\Delta g_{\mu\nu}$ will contain a shock wave: two wormhole solutions glued together along the surface normal to the trajectory given by $U-\gamma f(V)=0$. Such solutions were first constructed for a BH in \cite{thooft1,thooft2} and studied in detail in \cite{sfetsos}. This latter paper constructs the shock wave solution for a number of rather general metrics, however our wormhole does not fit into any of the classes considered there; it is therefore no surprise that a pure shock-wave solution does not exist in our case. We thus look for a solution containing a smooth correction of the shock wave. Following \cite{sfetsos}, the shock wave can be formulated (equivalently to the gluing picture) as a discontinuous coordinate change with an (as yet undetermined) discontinuity $c(U,V)$:
\bea
\nonumber (U,V)&\mapsto& (\tilde{U},\tilde{V})=\left(U,V-c\left(U,V\right)\Theta\left(U-\gamma f\left(V\right)\right)\right)\\
\nonumber (dU,dV)&\mapsto& (d\tilde{U},d\tilde{V})=\left(dU,dV-c\left(U,V\right)\tilde{\delta}dU+\gamma c(U,V)\tilde{\delta}f'(V)dV,dV\right)\\
\label{shockshift}\tilde{\delta}&\equiv&\delta\left(U-\gamma f\left(V\right)\right).
\eea
The last line is the equation of trajectory, already mentioned in relation to the geodesic equations. The above coordinate change influences all the tensors, in particular the metric $g_{\mu\nu}$ and the background stress-energy tensor $T_{\mu\nu}$. The metric correction from the background (\ref{metricbtz}) now totals the shock-wave contribution $\Delta g_{\mu\nu}$ plus the smooth contribution $\delta g_{\mu\nu}$:
\bea
\nonumber g_{\mu\nu}&\mapsto& g_{\mu\nu}+c(U,V)\Delta g_{\mu\nu}+\delta g_{\mu\nu}\\
\label{metricshock}\Delta g_{\mu\nu}&=&\left(\begin{matrix}-2g_{UV} & -\gamma\tilde{h}+2\gamma g_{UV}f'(V)+\gamma^2\tilde{h} & 0\\
                                                                             -\gamma\tilde{h}+2\gamma g_{UV}f'(V)+\gamma^2\tilde{h} & 2\gamma^2\tilde{h}f'(V) & ~~~0~~~\\
                                                                                                     ~~~0~~~ & ~~~0~~~ & ~~~0~~~\end{matrix}\right).
\eea
For symmetry reasons, the nonzero components of the smooth part are $UU$, $VV$, $UV$ and $\phi\phi$. Similar reasoning holds for the stress-energy tensor: being a second-rank tensor, $T_{\mu\nu}$ transforms the same way as the metric. Adding up $\delta T_{\mu\nu}$ and the direct contribution $t_{\mu\nu}$ from (\ref{tmunucuu}-\ref{tmunucuv}), the total stress-energy tensor is now
\bea
\nonumber T_{UU}&\mapsto& T_{UU}+t_{UU}-2c(U,V)\tilde{\delta}T_{UV}\\
\nonumber T_{UV}&\mapsto& T_{UV}+t_{UV}-c(U,V)\tilde{\delta}\gamma T_{VV}-c(U,V)\tilde{\delta}\gamma T_{UV}f'(V)+c(U,V)\tilde{\delta}\gamma^2T_{VV}\\
\label{tmunutot}T_{VV}&\mapsto& T_{VV}+t_{VV}+2c(U,V)\tilde{\delta}\gamma^2T_{VV}f'(V).  
\eea
Now, given a WH model, i.e. the function $T_{UU}(U)$, we can in principle write down and solve the Einstein equations. The unknown jump $c$ is found by matching the metric perturbation to its stress tensor and integrating across $U=f(V)$ to find the coefficient in front of the Dirac delta.


\emph{The fast wormhole shock waves.} Let us now do this explicitly for the fast wormhole. Consider first the regime $\Delta_1\geq 1/2$ with the Dirac delta metric (\ref{diracwhsol}), which is easier. The solution reads:
\bea
\label{diracwhc}c(U,V)&=&2p_0-\gamma^2p_0U_0\Theta(V-U_0)\frac{2V-1}{V}\\
\label{diracphph}\delta g_{\phi\phi}&=&\frac{2p_0^2}{1-p_0}g_{\phi\phi}\\
\label{diracuv}\delta g_{UV}&=&2\gamma p_0\frac{\mathrm{arctanh}(UV)}{(1+UV)^2},~~\delta g_{UU}=\delta g_{VV}=0.
\eea
Notice that (\ref{diracphph}) in fact has nothing to do with the wormhole (it is $\gamma$-independent), it is simply the higher-order correction to the linear shock wave. Also, this whole story rests on the small $\gamma$ expansion and thus is only valid when $UV\ll\gamma<1$, i.e. one should not worry that the hyperbolic arctangent grows exponentially for large $UV$. For $\Delta_1<1/2$ the algebra is more tedious; when the dust settles, $\delta g_{\phi\phi}$ is expectedly the same as in (\ref{diracphph}) while $c(U,V)$ and $\delta g_{UV}$ differ:
\bea
\label{crapwhc}c(U,V)&=&2p_0-2\gamma^2p_0\left(\frac{2\delta}{1-2\delta}\right)^2\left(1+\left(1-2\delta\right)\log\left(U-U_0\right)\right)U^{1-2\delta}\Theta(V-U_0)~~~~~~\\
\label{crapuv}\delta g_{UV}&=&\gamma p_0\exp\left(\frac{3-2\Delta_1}{2-4\Delta_1}\right)U^{3/2+\Delta_1}(1+(1-2\Delta_1)\log U).
\eea
This completes the solution for the wormhole geometry perturbed by the eikonal scalar waves. The primary qualitative feature is the deformation of the shock wave, i.e. the wavefront has the form determined by the trajectory equation, and the amplitude is likewise spacetime-dependent. This effectively introduces long-time and nonlocal correlations that can kill the fast scrambling.

Now we can put the pieces together to express the on-shell action. We start from the textbook linearized gravity-matter action (\ref{sclass}). For our backreacted metric (\ref{metricshock}), with both shock wave and non-shock wave contribution, it becomes:
\be
\label{linact}S_c=\frac{1}{4}\int d^3x\sqrt{-g}\left(c\Delta g_{\mu\nu}+\delta g_{\mu\nu}\right)t^{\mu\nu}.
\ee
Note that the matter-independent kinetic term (of the form $\Delta g_{\mu\nu}\partial^2\Delta g^{\mu\nu}$) is in fact included in (\ref{linact}) as it equals minus one half of the metric-matter terms of the form $\Delta g_{\mu\nu}t^{\mu\nu}/2$. The Dirac delta terms, coming from the shock wave, will only contribute along the line $U=\gamma f(V)$, but the smooth terms will contribute to the integral in (\ref{linact}) in the whole space. Finally, on top of the waves $\Psi_{1,3}$ with asymptotic momentum $p$, we add up also the waves $\Psi_{2,4}$ with asymptotic momentum $q$, which are easily obtained by symmetry from the solutions already found.

As usual, the Dirac delta regime of the fast wormhole is the simpler case. Since both the shock wave and the bare wormhole metric contain a Dirac delta, the phase contains terms linear and quadratic in Dirac deltas:
\bea
&&S_c^{(\Delta_1\geq 1/2)}=\int_{U_0}^\infty du\int_{U_0}^\infty dv\Bigg[\left(-\frac{3}{8}\gamma^2U_0^2\frac{1-uv}{1+uv}\left(p_0q_0\frac{1-u^2(4v+v^2-4U_0)}{u^2v^2}+p_0^2\frac{1-v^2(4v+u^2-4U_0)}{v^4}\right)\right)\delta(u)+\nonumber\\
&&+\left(-\frac{3}{8}\gamma^2U_0^2\frac{1-uv}{1+uv}\left(q_0^2\frac{1-u^2(4v+v^2-4U_0)}{u^2v^2}+p_0q_0\frac{1-v^2(4v+u^2-4U_0)}{u^4}\right)\right)\delta(v)\nonumber\\
&&+p_0q_0\left(1-4\gamma U_0\right)\delta(u)\delta(v)+\left(\ldots\right)\delta(u)^2+\left(\ldots\right)\delta(v)^2\Bigg].\label{linact0dirac}
\eea
The range of the integrals comes from integrating over the Dirac deltas in solving for the perturbed metric, i.e. from the Heaviside step functions in the solution for $c(U,V)$ in earlier equations. We have not written explicitly the terms in front of the squares of Dirac deltas as these give zero under the integral, according to the usual Colombeau algebra or the physical arguments in \cite{sfetsos}. The $\delta(u)\delta(v)$ term in the last line of (\ref{linact0dirac}) gives no difficulties, however the first line requires us to expand around $u=0$ (direct insertion of $u=0$ yields an infinity), then integrate over $v$ from $U_0<0$ to $\Lambda>0$ taking the principal value at $v=0$ (otherwise we end up with divergences), and finally take the limit $\Lambda\to\infty$. Analogous steps hold for the second line in (\ref{linact0dirac}), just replacing the $u$ and $v$ integration. Denoting the first line of (\ref{linact0dirac}) by $I_1$, we get
\bea 
I_1&=&-\lim_{u_0\to 0}\frac{3}{8}\gamma^2U_0^2\int_{-U_0}^\Lambda\left(\frac{1+4U_0v^2-v^3}{v^4}p_0^2+\frac{1}{u_0^2v^2}p_0q_0\right)=\nonumber\\
&=&-\lim_{u_0\to 0}\lim_{\epsilon\to 0}\lim_{\Lambda\to\infty}\frac{3}{8}\gamma^2U_0^2\left(\int_{-U_0}^{-\epsilon}+\int_\epsilon^\Lambda\right)\left(\frac{1+4U_0v^2-v^3}{v^4}p_0^2-\frac{p_0q_0}{u_0^2\Lambda}\right)=\nonumber\\
&=&\frac{\gamma^2}{8U_0}p_0^2\left(1+12U_0^3\left(1+\log\frac{\Lambda}{U_0}\right)\right)+O\left(\frac{1}{\Lambda}\right).\label{linact0int}
\eea
We simply ignore the divergent contribution $\propto\log\Lambda$ coming from the $v\to\infty$ region -- since we are only interested in the phase shift we do not care about the constant infinite term, which anyway clearly comes from long-time (far infrared) processes which as a rule require regularization in scattering problems. Of course, the terms proportional to $1/\Lambda$ go to zero so they are also ignored. Finally, summing the value of (\ref{linact0int}), the value of the analogous integral for the $\delta(v)$ term and the (simple) integral over the $\delta(u)\delta(v)$ term in (\ref{linact0dirac}), we obtain for the phase shift:
\be
\label{linactdirac}S_c^{(\Delta_1\geq 1/2)}=p_0q_0(1-4\gamma U_0)-\frac{\gamma^2}{8U_0}\left(1+12U_0^3\right)\left(p_0^2+q_0^2\right).
\ee
This can be compared to the black hole result $S_c=pq=p_0q_0$:\footnote{We pick the units so that $4\pi G=1$.} the main effect is the appearance of terms $p_0^2$ and $q_0^2$.

In the regime $\Delta_1<1/2$ the calculations are similar except that the wormhole metric contains no shock waves, hence we only get terms with $\delta(u)$, $\delta(v)$ and $\delta(u)\delta(v)$ but not the terms with squares of delta functions. But the latter are irrelevant anyway, hence the calculation closely follows the sequence (\ref{linact0dirac}-\ref{linactdirac}). We thus get:
\bea
\nonumber S_c^{(\Delta_1<1/2)}&=&p_0q_0(1-4\gamma U_0)-\gamma\left(\frac{\Delta_1}{1-2\Delta_1}\right)^2\exp\left(\frac{3-2\Delta_1}{2-4\Delta_1}\right)\left(p_0^2-q_0^2\right)-\\
\label{linactcrap}&&-14\gamma^2\left(\frac{\Delta_1}{1-2\Delta_1}\right)^4\left(p_0^2+q_0^2\right).
\eea
Again, in addition to the center-of-mass momentum squared $p_0q_0$, we have also the $p_0^2$ and $q_0^2$ terms. 

\subsubsection{The scattering amplitude}

Now we can calculate the scattering amplitude in the time-dependent, horizonless wormhole background. Such situations are considered in the seminal work \cite{balasubra2019otoc} where the 
authors rewrite the amplitude in the coordinate representation in time-ordered form with the help of the Green identity. This is a more elegant and physically transparent way than what we do here, however in the wormhole geometry we have found it difficult to find the boundary surfaces along which to integrate the Green identity, so we have not succeeded in the time-ordered method here. Instead, we will rewrite (\ref{otocint}) in terms of coordinates and nonconserved momenta with explicit proper-time dependence, and then we will again make a perturbative expansion of the momenta around their asymptotic values. 

The general formula (\ref{otoc}) remains valid. But the following differences arise: (1) the infalling geodesic is now given by (\ref{eqorbit}) as $U=\gamma f(V)$ (2) the momentum $p$ receives corrections and does not stay equal to $p_0$ for all times. The resulting wavefunctions read (compare to (\ref{otocfuns1}-\ref{otocfuns2})):
\bea
\Psi_1(t;p)&=&\int dU\int dVe^{\imath\frac{g_{UV}}{2}\left(p^VU+p^UV\right)}K_1(U,V;t)\delta\left(U-\gamma f\left(V\right)\right)\nonumber\\
\Psi_2(0;q)&=&\int dU\int dVe^{\imath\frac{g_{UV}}{2}\left(q^VU+q^UV\right)}K_2(U,V;0)\delta\left(V-\gamma f\left(U\right)\right).~~~\label{otocfunswh}
\eea
In principle, we could still easily get rid of one half of coordinate integrations thanks to the Dirac deltas. However, the resulting expressions are intractable unless we expand the geodesic and the metric in $\gamma$ small; this is justified as the geodesic equation and the metric were themselves found as expansions in $\gamma$. On the other hand, we do not expand the \emph{propagators} $K_{1,2}$ themselves as functions of $\gamma$, as their $\gamma$-dependence is nonperturbative, obtained by summing over all the modes. This means we first expand the Dirac deltas as $\delta(U-\gamma f(V))=\delta(U)-\gamma f(V)\delta'(U)+\ldots$, leading to:
\bea
\Psi_1(t;p)&=&\int dU\int dVe^{\imath\frac{g_{UV}}{2}\left(p^VU+p^UV\right)}K_1(U,V;t)\left(\delta\left(U\right)-\gamma f\left(V\right)\delta'\left(U\right)\right)\nonumber\\
\Psi_2(0;q)&=&\int dU\int dVe^{\imath\frac{g_{UV}}{2}\left(q^VU+q^UV\right)}K_2(U,V;0)\left(\delta\left(V\right)-\gamma f\left(U\right)\delta'\left(V\right)\right).~~~\label{otocfunswhexp}
\eea
Then we also expand the metric component $g_{UV}$ in the exponents in (\ref{otocfunswh}). This yields the following structure of the amplitude:
\bea
D(t,0)&=&\int p_0dp_0\int q_0dq_0e^{\imath S_c(p_0,q_0)}\mathcal{A}=\int p_0dp_0\int q_0dq_0e^{\imath S_c(p_0,q_0)}\left(\mathcal{A}_0+\gamma\mathcal{A}_1+\ldots\right)\nonumber\\
\mathcal{A}_0&=&k_1^\dagger k_2^\dagger k_1k_2\nonumber\\
\mathcal{A}_1&=&\mathcal{K}_1^\dagger k_2^\dagger k_1k_2+k_1^\dagger\mathcal{K}_2^\dagger k_1k_2+k_1^\dagger k_2^\dagger\mathcal{K}_1k_2+k_1^\dagger k_2^\dagger k_1\mathcal{K}_2+
\int 2\imath K_1^\dagger\left(0,V;t\right)K_2^\dagger\left(U,0;0\right)K_1\left(0,V';t\right)K_2\left(U',0;0\right)\times\nonumber\\
&\times&\left[p_0\left(U^2f(U)-U'^2f(U')-h(U,0)+h(U'0)\right)+q_0\left(V^2f(V)-V'^2f(V')-h(V,0)+h(V',0)\right)\right]\label{inoutuv0}
\eea
where for brevity we write the Fourier transforms of propagators and their derivatives as
\bea
k_1&\equiv&\int dVe^{\imath\frac{g_{UV}}{2}p_0V}K_1\left(0,V;t\right),~~k_2\equiv\int dUe^{\imath\frac{g_{UV}}{2}q_0U}K_2\left(U,0;0\right)\nonumber\\
\mathcal{K}_1&\equiv&\int dVe^{\imath\frac{g_{UV}}{2}p_0V}f(V)\partial_VK_1\left(0,V;t\right),~~\mathcal{K}_2\equiv\int dUe^{\imath\frac{g_{UV}}{2}q_0U}f(U)\partial_UK_2\left(U,0;0\right)\nonumber\\
\label{k12}
\eea
The phase, i.e. the classical on-shell action is also different from the BH geometry; it is given in (\ref{linactdirac}) or (\ref{linactcrap}) depending on the WH model. The propagators $K(U,V;\phi)$ in the WH background were found in the subsection \ref{sec2kg}. The expressions (\ref{inoutuv0}-\ref{k12}) suggest how to proceed with the practical calculations: we can first find the $U$/$V$ integrals of propagators and their derivatives $k_1,k_2,\mathcal{K}_1,\mathcal{K}_2$ in
(\ref{k12}); the last term in (\ref{inoutuv0}) is the only one which has more than a single coordinate integral but it still not too difficult. When all coordinate integrations are done we can insert the resulting expressions which now depend solely on $p_0,q_0$ into the main $\int dp_0\int dq_0$ integral in (\ref{inoutuv0}) and solve it in the saddle-point approximation similar to the method of \cite{butterstring}.\footnote{An alternative path to the amplitude calculation is to work in the shock wave background, when there is no scattering but the transformed wave functions in the new geometry give rise to a multiplicative factor in the amplitude, which coincides with the phase $\exp(\imath S_c)$. In the first draft of this paper we have combined the two approaches but we have overcounted the phase. In the current version we have corrected this error and have done all the calculations in the WH frame which turns out to be simpler. The qualitative conclusions do not change but the exact values of the Lyapunov exponents do differ.}

This is the final outcome of our formalism for OTOC calculation. The rest is just algebra (actually, elementary integrals, saddle-point integration and transformations with hypergeometric functions), but the outcome of this algebra is the core of the paper -- the behavior of OTOC and the Lyapunov exponents. We devote the next section to a detailed discussion of these matters.

\section{Lyapunov spectra for fast wormholes}\label{sec4}

Now we will describe the behavior of OTOC in various parameter regimes. The relevant variables are the conformal dimensions (bulk masses) $\Delta_{1,2}$ and the wormhole coupling $\gamma$. The result 
is the Lyapunov spectrum, the collection of exponents $\lambda$ which characterize the correlation decay. We compute the OTOC integral (\ref{inoutuv0}) for the fast wormhole model, and then we plot the spectrum of Lyapunov exponents for various cases and discuss the physical consequences. In \ref{appc} we compare the results to the more difficult case of a long-living (slow) wormhole.

Before we take off, two technical remarks are in order. First, all results for OTOC are of course time-dependent functions multiplied by time-independent constants depending on $\Delta_{1,2}$, $\gamma$ and $r_h$. These constant terms are not important for us as we are mainly interested in the time dependence, not the absolute magnitude of the function $D(t,0)$. For this reason we always just leave out such constant terms. Second, we will emphasize the essence over the calculational details; therefore we sometimes leave out the full integral as calculated in the saddle-point approximation (when the expression is unpractically long) and give only the asymptotic long-time dynamics in terms of exponentials or power laws.

Consider first the simpler case, the Dirac delta regime ($\Delta_1>1/2$) of the fast wormhole. Plugging in the propagator (\ref{kbnddirac}) and feeding the function $f$ from  (\ref{eqorbit}), the coordinate integrals in (\ref{k12}) can all be performed exactly. For $k_{1,2}$ the outcome is: 
\bea
k_1&=&\frac{2\pi}{\Gamma(\Delta_1)}e^{-\frac{\imath\pi\Delta_1}{2}+\frac{\gamma U_0}{4}}e^{\imath e^{r_ht}\tilde{p}}\left(e^{r_ht}\tilde{p}\right)^{\Delta_1-1}\nonumber\\
k_2&=&\frac{2\pi}{\Gamma(\Delta_2)}e^{-\frac{\imath\pi\Delta_2}{2}+\frac{\gamma U_0}{4}}e^{\imath\tilde{q}}\tilde{q}^{\Delta_2-1}\label{k12fast}
\eea
where we introduce the rescaled asymptotic momenta
\be
\label{tildepq}\tilde{p}=p_0r_h^{\gamma U_0/2}e^{\frac{\gamma U_0}{4}(\log 4-1)},~~\tilde{q}=q_0r_h^{-\gamma U_0/2}e^{-\frac{\gamma U_0}{4}(\log 4-1)}.
\ee
The integrals $\mathcal{K}_{1,2}$ are also obtained analytically, however the outcome is very complicated; we will give the final expressions for the OTOCs which are actually simpler.

Inserting $k_{1,2},\mathcal{K}_{1,2}$ into (\ref{inoutuv0}), we find that the integral of $\mathcal{A}_0$ is not problematic but the integral of $\mathcal{A}_1$ is intractable (even in the saddle-point approximation) unless we expand $\mathcal{K}_{1,2}$ in $\gamma$. This yields the following momentum integral (up to $O(\gamma^2)$ terms):  
\bea 
&&D^{(\Delta\geq 1/2)}(t,0)=\frac{(2\pi)^4}{\Gamma^2(\Delta_1)\Gamma^2(\Delta_2)}\int d\tilde{q}\tilde{q}^{2\Delta_1-1}\int d\tilde{p}\left(e^{r_ht}\tilde{p}\right)^{2\Delta_2-1}e^{-\imath e^{r_ht}\tilde{p}-\imath\tilde{q}}\times~~~~~\nonumber\\
&&\times e^{\imath\tilde{p}\tilde{q}(1-4\gamma U_0)+\frac{3}{8}\imath\gamma^2/U_0\left(1+12U_0^3\right)(\tilde{p}^2+\tilde{q}^2)}\left[1+\gamma r_h^{-\gamma\Delta_2U_0}e^{-\frac{\gamma\Delta_2U_0}{2}(\log 4-1)}e^{(2\Delta_1-1)r_ht}\tilde{p}^{2\Delta_1}\tilde{q}e^{2\imath\tilde{p}e^{r_ht}+2\imath\tilde{q}}\right]~~~~~~~\label{otocintdirac}
\eea
The integral over momenta in (\ref{otocintdirac}) is doable in the saddle-point approximation, as in \cite{butterstring}. The saddle-point approximation works best when one conformal dimension is significantly larger than the other. We must therefore distinguish the cases $\Delta_2>\Delta_1$ and $\Delta_1>\Delta_2$. The two are not symmetric because the operators with $\Delta_1$ are inserted at time $t$ and those with $\Delta_2$ at time $0$, and indeed the Lyapunov spectra will differ.

\subsection{Lyapunov spectrum with $\Delta_1>\Delta_2,1/2$}

Assume first that $\Delta_1$ is larger. The saddle point corresponds simply to $\tilde{q}_*\approx\Delta_1$; inserting this in the remaining $\tilde{p}$-integral in (\ref{otocintdirac}) and shifting the contour as $t\mapsto t-\imath\epsilon$ yielding a Jeans-type integral in $q$, which solves in terms of hypergeometric functions. The important point is the asymptotic behavior, obtained for small $\gamma$ and large $t$:
\bea 
&&D^{(\Delta_1>\Delta_2)}(t,0)\sim\frac{1+e^{r_ht}}{\left(1+\frac{2}{\Delta_1}e^{r_ht}\right)^{2\Delta_2+1}}+\nonumber\\
&&+\gamma^{4\Delta_2-1}\exp\left(\frac{2\imath}{\gamma^2}\left(1+\frac{2}{\Delta_1}e^{r_ht}\right)^2\right)\left[1+e^{r_ht}+\frac{\pi^2}{\Gamma(\Delta_1)^2}\left(1+\frac{2}{\Delta_1}e^{r_ht}\right)^{2\Delta_2-1}\right]^{-1}.~~~~~~\label{otocdirac1asymp}
\eea
Notice that the first term is (up to multiplicative constants) identical to the black hole results from \cite{butter,butterstring}; the second term is novel. It contains a rapidly oscillating factor, its frequency diverging as $1/\gamma^2$. However there is no reason to worry about the $\gamma\to 0$ limit since the whole second term vanishes in that limit as $\gamma^{4\Delta_2-1}$. The oscillations are not that surprising: in a wormhole, there is no infalling boundary condition as for a black hole; the analytical solution that we have constructed generically mixes the infalling and outgoing modes, hence the oscillatory contribution to the correlator. The amplitude of the oscillations is in fact relatively small for typical parameter values, they constitute a minor modulation of the dominant black-hole-like result. Even though the time dependence of (\ref{otocdirac1asymp}) is quite complicated, we can put everywhere $t\mapsto t/r_h$ and get rid of all the exponents, hence the Lyapunov exponent is still:
\be 
\lambda^{(\Delta_1>\Delta_2)}=2\pi T.
\ee
This will change in the other regime.

\subsection{Lyapunov spectrum with $\Delta_2>\Delta_1>1/2$}

\begin{figure}
\includegraphics[width=.48\textwidth]{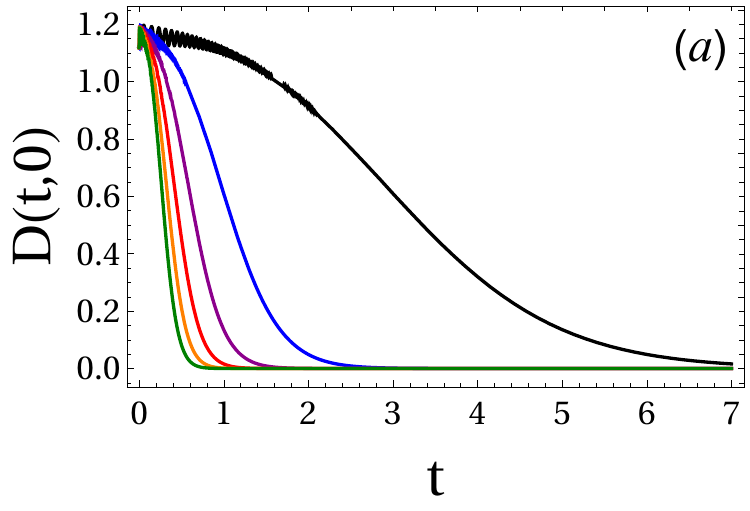}
\includegraphics[width=.48\textwidth]{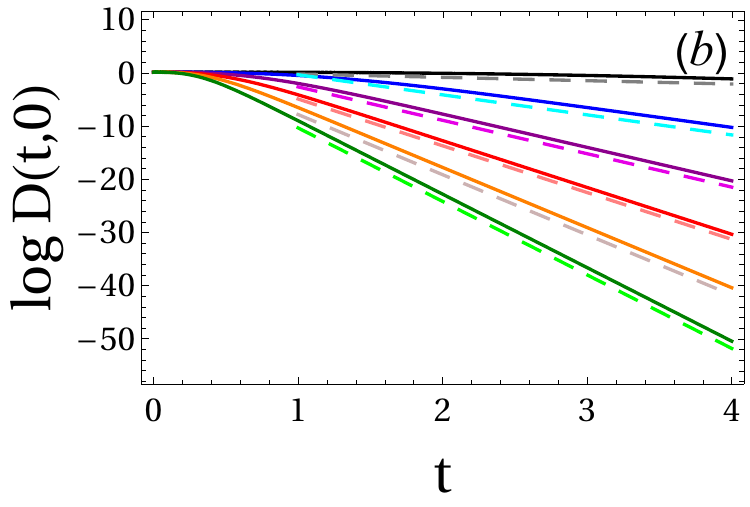}
\includegraphics[width=.48\textwidth]{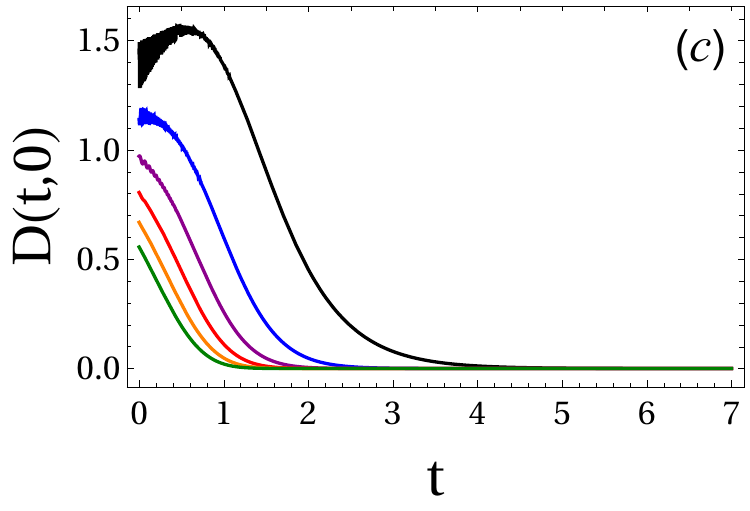}
\includegraphics[width=.48\textwidth]{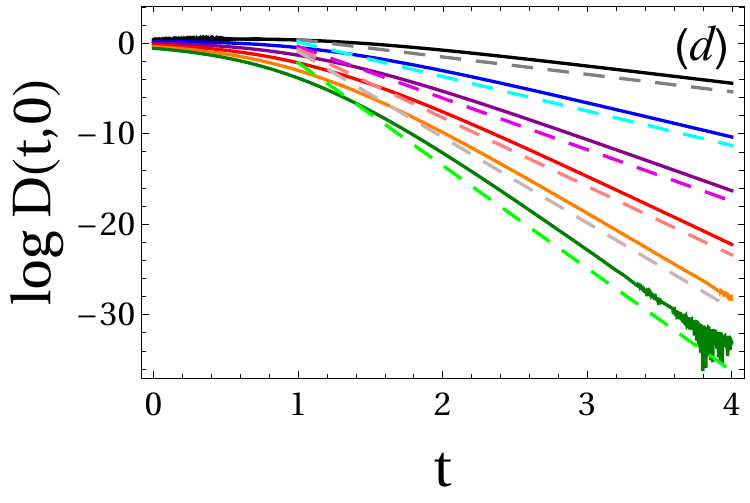}
\caption{The out-of-time-ordered correlator $D(t,0)$ for the case $\Delta_2>\Delta_1>1/2$, as a function of temperature (top row -- a,b) and the conformal dimension $\Delta_1$ (bottom row -- c,d); the left-hand side panels (a,c) show the OTOC value on the linear scale and the right-hand side panels (b,d) show exactly the same functions on the logrithmic scale. In (a,b) the temperatures are $T=0.1,0.3,0.5,0.7,0.9,1.1$ (black, blue, violet, red, yellow, green), with $\Delta_1=1$, whereas in (c,d) the conformal dimension is varied as $\Delta_1=0.51,1.0,1.5,2.0,2.5,3.0$ (black, blue, violet, red, yellow, green) with $T=0.3$; notice that we have shifted the first value above the threshold dimension $\Delta_1=0.5$. In all plots we have $\Delta_2=10$ and $\gamma=0.3$. In (b) we plot also the asymptotic curves $\exp(-4\Delta_1\pi Tt)$ for comparison -- at longer times all curves precisely follow the universal falloff in accordance with the MSS bound but on shorter scales the behavior is very non-universal which is particularly obvious on the linear scale in (a). In (d) we see the same long-time law $\exp(-4\Delta_1\pi Tt)$, again with significant additional features on shorter timescales.}
\label{figotoc2}
\end{figure}

In this case there is no simple saddle point in $\tilde{q}$ but we can do the saddle-point approximation in $\tilde{p}$, yielding $\tilde{p}_*\approx\Delta_2e^{-r_ht}$. The integral over $\tilde{q}$ is again reduced to Gaussian and Jeans integrals. Expanding again for small wormholes and long times, we obtain
\bea 
&&D^{(\Delta_2>\Delta_1>1/2)}(t,0)\sim\frac{1+e^{r_ht}}{\left(1+\frac{2}{\Delta_2}e^{r_ht}\right)^{2\Delta_1+1}}+\nonumber\\
&&+\gamma^{4\Delta_1-1}\exp\left(-\frac{2\imath}{\gamma^2}\left(1+\frac{2}{\Delta_1}e^{r_ht}\right)^2\right)\left(1+e^{r_ht}+\frac{\Delta_2}{\Delta_1^2+2\Delta_1+3}e^{2r_ht}\right)^{-1}\times\nonumber\\
&&\times\left(e^{-\frac{2\Delta_1}{4\Delta_1-1}r_ht}+\frac{2}{\Delta_2}e^{-\frac{1}{4\Delta_1-1}r_ht}\right)^{2\Delta_1-1}.\label{otocdirac2asymp}
\eea
This function is even more contrived than (\ref{otocdirac2asymp}) but crucial is the fact that we cannot introduce a single scaling exponent: putting $t\mapsto t/r_h$ would still leave nontrivial exponents in the last line of (\ref{otocdirac2asymp}). Now we have to introduce the vector of Lyapunov exponents $\vec{\lambda}$ which in this case has three components, the first saturating the chaos bound but the other two being lower:
\be 
\vec{\lambda}^{(\Delta_2>\Delta_1>1/2)}=\left(2\pi T,\frac{2\Delta_1}{4\Delta_1-1}2\pi T,\frac{1}{4\Delta_1-1}2\pi T\right).\label{lambda2asymp}
\ee
One might worry that only the largest exponent makes sense, i.e. the growth is as fast as its fastest mode. Indeed, in classical dynamics one also typically computes the largest Lyapunov exponent. But this is not the whole story and a complete characterization of chaos requires the whole Lyapunov spectrum. We simply do not know which Lyapunov mode is dominant in a certain time window without plotting the OTOC for specific parameter values. In Fig. \ref{figotoc2} we plot the function (\ref{otocdirac2asymp}) for a few values of conformal dimensions and the temperature. The panels (a,b) show the temperature dependence on linear and logarithmic scales, respectively -- on longer timescales all curves precisely follow the law $\exp(-2\Delta_1\times 2\pi Tt)$, saturating the MSS bound\footnote{One should not be confused about the extra $2\Delta_1$ factor, which is present also in the analytic form (\ref{otocdirac2asymp}) -- it is just the overall dimensional factor which does not enter the scaling of $t$ with $\lambda$ as it does not directly multiply the time-dependent part.} but there is also the early regime, before all the curves collapse to the fastest (MSS) exponential decay, where an interplay of all exponents in $\vec{\lambda}$ is seen. For low temperatures and low $\Delta_1$ values, the oscillatory prefactor is also seen at low times.

\subsection{Lyapunov spectrum with $\Delta_2>1/2>\Delta_1$}

So far we have focused on the simpler wormhole solution obtained when $\Delta_1>1/2$. In the regime when $\Delta_1<1/2$, the logic is similar as before except that we can only have the case $\Delta_2>\Delta_1$. The expressions for the momentum integrals become more complicated but the asymptotic behavior is not too different from (\ref{otocdirac2asymp}):
\bea 
&&D^{(\Delta_2>1/2>\Delta_1)}(t,0)\sim\frac{1+e^{r_ht}}{\left(1+\frac{2}{\Delta_2}e^{r_ht}\right)^{2\Delta_1+1}}+\nonumber\\
&&+\gamma^{2\Delta_1+1/2}\exp\left(-\frac{\sqrt{6}\imath}{\gamma^{4\Delta_1}}\left(1+\frac{2}{\Delta_1}e^{r_ht}\right)^2\right)\left(1-U_0+e^{r_ht}+\frac{\Delta_2}{\Delta_1^2+1}e^{2r_ht}\right)\times\nonumber\\
&&\times\left(e^{-\frac{\Delta_1^2-5\Delta_1+3}{2\Delta_1+4}r_ht}+\frac{2}{\Delta_2}e^{-\frac{\Delta_1}{\Delta_2}r_ht}\right)^{2\Delta_1-1}.\label{otocdirac3asymp}
\eea
There is again a spectrum of three exponents, two of them below the chaos bound:
\be 
\vec{\lambda}^{(\Delta_2>1/2>\Delta_1)}=\left(2\pi T,\frac{\Delta_1^2-5\Delta_1+3}{2\Delta_1+4}2\pi T,\frac{\Delta_1}{\Delta_2}2\pi T\right).\label{otocspec3}
\ee
For an easier grasp of the behavior of the result (\ref{otocdirac3asymp}), we again plot the function $D(t,0)$ for a few sets of parameter values. Just as in the previous case, a generic behavior is a complicated function of different exponentials, which only approximately correspond to (\ref{otocspec3}) at any given moment, but in general at longer times the MSS bound is approached. Importantly, the early regime is in fact more representative of chaos -- at longer times, when saturation is approached, the exponential scrambling gives way to more complex phenomena such as Ruelle resonances. 

\begin{figure}
\includegraphics[width=.48\textwidth]{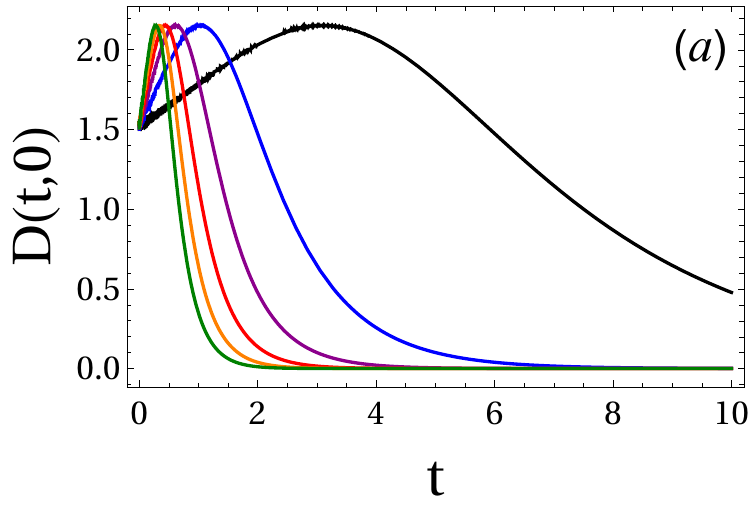}
\includegraphics[width=.48\textwidth]{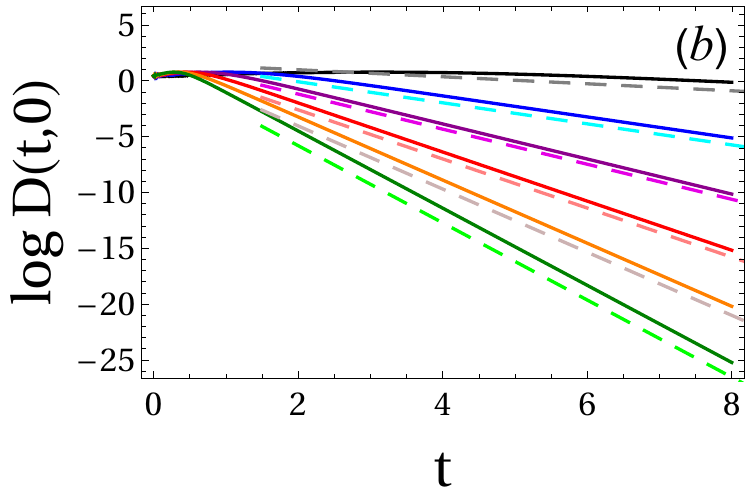}
\includegraphics[width=.48\textwidth]{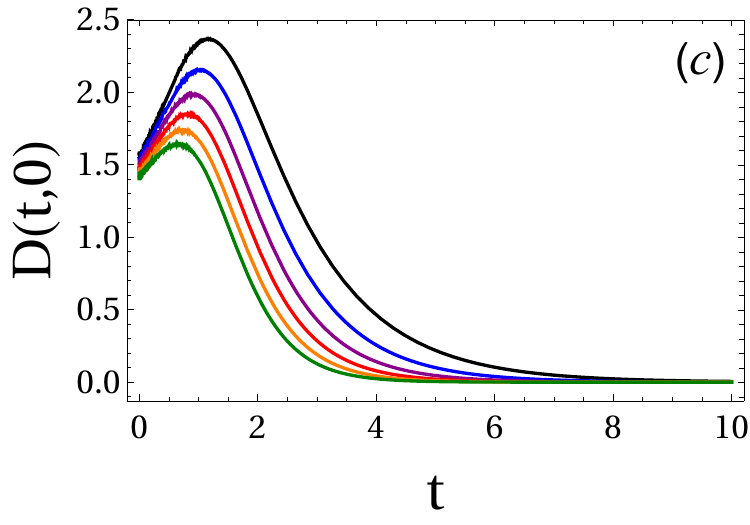}
\includegraphics[width=.48\textwidth]{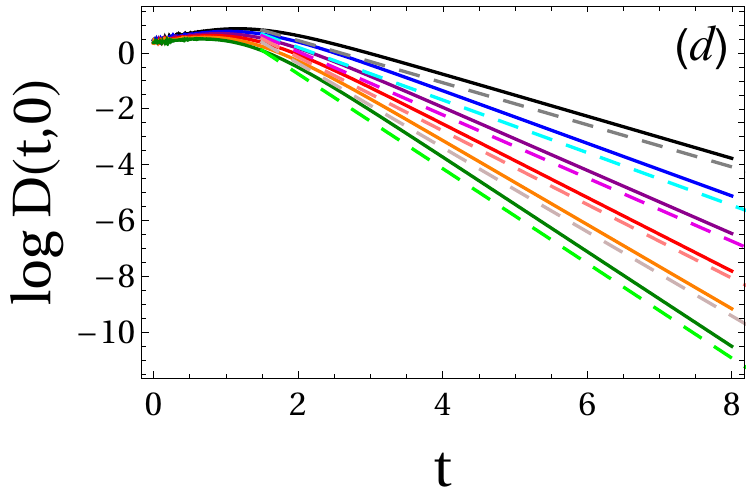}
\caption{The OTOC value $D(t,0)$ for the case $\Delta_2>1/2>\Delta_1$, as a function of temperature (top row -- a,b) and the conformal dimension $\Delta_2$ (bottom row -- c,d); the left-hand side panels (a,c) show the OTOC value on the linear scale and the right-hand side panels (b,d) show exactly the same functions on the logarithmic scale. In (a,b) the temperatures are $T=0.1,0.3,0.5,0.7,0.9,1.1$ (black, blue, violet, red, yellow, green), with $\Delta_1=0.25$, whereas in (c,d) the conformal dimension is varied as $\Delta_1=0.20,0.25,0.30,0.35,0.40,0.45$ (black, blue, violet, red, yellow, green) with $T=0.3$; for all plots we have $\Delta_2=10$ and $\gamma=0.3$. In (b,d) we plot also the asymptotic curves $\exp(-4\Delta_1\pi Tt)$ for comparison -- again the non-universal and $\Delta_1$-dependent behavior is seen at early times.}
\label{figotoc3}
\end{figure}

\subsection{Resume}

Let us collect and systematize the findings obtained in the previous subsections. This essentially amounts to a phase diagram of wormholes in terms of scalar perturbations.
\begin{itemize}
 \item{\emph{Maximum chaos.}} Fast wormholes always have a maximally chaotic mode, which retains the fast black hole scrambling mechanism with the exponent $2\pi T$. This is the consequence of the exponential boost factor $e^{r_ht}$ at the horizon. Even though a wormhole opens at time $t_0$, there will always be orbits launched early enough from the AdS boundary that do not see the wormhole throat.
 \item{\emph{Fast chaos.}} In the regime $\Delta_2>\Delta_1$ the fast wormhole Lyapunov exponent falls below the maximum, but is still linear in temperature, i.e. the BH-like boost is still present. In addition, the submaximum chaos only influences the short timescales -- the long-time approach to saturation is always dominated by maximal chaos.
 \item{\emph{Slow chaos or zero chaos.}} As we show in \ref{appc}, there are strong numerical indications that for a "slow" wormhole, i.e. a long-living wormhole where the wormhole lifetime is the largest scale in the system, one can find an exponentially small or even zero Lyapunov exponent. The intuitive explanation is that slow wormholes are created in distant past, so almost all orbits see the wormhole mouth. No wonder this regime has no linear scaling of $\lambda$ with temperature: there is no exponential boost factor characteristic for the black hole. These are however only phenomenological claims, as we do not have a good analytic control over this case. This is also the reason that the slow wormhole calculation is delegated to an Appendix.
\end{itemize}

\section{Discussion and conclusions}\label{sec5}

The basic findings, summed up in the previous subsection, are in a sense expected: we know from \cite{scramble,scramble2} that black holes are the fastest scramblers,\footnote{From a rigorous viewpoint this is actually still a conjecture, but physical arguments accumulated in the meantime support it strongly.} thus it is no wonder that removing the black hole horizon slows down the scrambling. Since fast wormholes still have a horizon some of the time, it is also logical that their Lyapunov spectrum always contains one exponent with the MSS value $2\pi T$, together with other, smaller exponents which are only important in the transient regime, long before saturation. Most of the time, OTOC is dominated by the fastest, MSS chaos rate.

The difficult case of slow wormholes is for now too demanding for analytical work. The numerical results from \ref{appc} suggest a scaling of the Lyapunov exponent of the form $\exp(-\mathrm{const.}/T)$. A work on the exact eternal SYK wormhole from field theory side \cite{otocsykexp} has found \emph{nonzero Lyapunov exponent with the same type of scaling} $\exp(-E_g/2T)$ where $E_g$ is the energy gap in the wormhole phase. Is it more than a coincidence? Maybe yes because (1) it might be a universal effect independent of the details of the model (2) slow wormholes should have a smooth limit to eternal ones. Maybe no because (1) the eternal SYK wormhole is a $0+1$-dimensional field theory, dual to $1+1$-dimensional Jackiw-Teitelboim (JT) gravity, very different from gravity in $2+1$ dimension that we consider (2) it is unlikely that the energy gap can be matched to the constant of proportionality that we find. We hope to learn more on this question in the future.

We have found that a wormhole in general has a full Lyapunov spectrum, i.e. a vector of Lyapunov exponents, rather than just a single exponent. This is expected for any system with many degrees of freedom, and likely holds for any model more complicated than a classical black hole. For example, generalizations of the SYK model such as \cite{altmanfl,altmanphtr,otocsach} likely also have a Lyapunov spectrum with several different exponents. The story of direction- and velocity-dependent Lyapunov exponents in \cite{otocsarosi,otocsarosi2} essentially means that systems with non-maximal chaos have a continuous Lyapunov spectrum; in our case, for spherical perturbations of wormholes, the spectrum is discrete (but would likely become continuous for localized shocks).

An interesting work for us is \cite{sykrec}: it considers a wormhole, again from field theory side, as two coupled SYK models dual to the eternal JT wormhole. The authors find recurrences (multiple peaks) in transmission probability for the wormhole, qualitatively similar to the oscillations in some of our plots; quite likely our OTOC sees these same recurrent states.


One might wonder how we could ever obtain anything but the simple MSS bound result $\lambda=2\pi T$ since we work in the classical gravity regime. The resolution is that opening a wormhole already requires quantum effects in the matter sector in order to break the null energy condition. It is true that gravity itself stays classical, however the removal of the horizon (which is basically what the GJW protocol does) seems enough to introduce scalings other than MSS. The fact that one exponent always stays at $2\pi T$ for any fast wormhole is certainly a consequence of the fact that for early enough times the horizon is always present. Another way to intuitively understand the appearance of exponents below $2\pi T$ is that the existence of both incoming and outgoing solutions results in an eikonal phase where the terms $p^2,pq,q^2$ all appear, rather than just the $pq$ term as for a black hole. 

In the future we plan to deal specifically with one aspect of our findings, related to the wormhole teleportation protocols; out of several proposed scenarios, we are mainly inspired by \cite{teleworm}. That paper develops in detail the idea from \cite{erepr2} where the left and right AdS boundary, i.e. field theory serve as entanglement resources, and we want to teleport the state of the qubit Q, inserted in the left CFT, to the qubit T inserted in the right CFT, with reference R which starts out maximally entangled with Q. Translated into correlation functions, the correlation between Q and T is just the expectation value of the anticommutator (because the operators are fermionic) of the two inserted observables (on the left and on the right) in the presence of double-trace coupling, which can be expressed in terms of TOC (which quickly factorizes and reaches a constant value) and OTOC.

Finally, another important goal for future work is to consider the quantum chaos and teleportation in more realistic strongly correlated systems, in particular the holographic non-Fermi liquids and strange metals. Once the appropriate  background is constructed, our perturbative formalism for the scattering amplitude can be readily applied. This might provide experimentally relevant predictions for the equilibration of the system.

\section*{Acknowledgments}

I am grateful to P.~Sabella-Garnier, K.~Schalm, J.~van~Gorsel, K.~Nguyen, R.~Espindola, V.~Jahnke, J.~Pedraza and Z.~Yang for helpful discussions. This work has made use of the excellent Sci-Hub service. Work at the Institute of Physics is funded by the Ministry of Education, Science and Technological Development and by the Science Fund of the Republic of Serbia, under the Key2SM project (PROMIS program, Grant No. 6066160).

\appendix

\section{Metrics and Klein-Gordon equation for the fast wormhole with $\Delta<1/2$}\label{appa}

Here we complement the section 2.2 with the metric in $(t,r,\phi)$ coordinates, solutions to the Klein-Gordon equation and the bulk-to-boundary propagators for the cases not given in the main text. We always divide the spacetime in the way given in the main text: the throat region and the far region in $r$, and the three regimes in time. Consider first the quick wormhole for  $\Delta>1/2$. The metric in the throat region reads
\be
\label{metricnearapp}ds^2=(1+\gamma^2r_h^2\rho^2)(-dt^2+d\phi^2)+\frac{d\rho^2}{1+\gamma^2r_h^2\rho^2}+\sigma\frac{\gamma Be^{(2\Delta_1-1)r_ht}\rho^{1-\Delta}}{1+\gamma^2r_h^2\rho^2}dtd\rho,
\ee
with $\sigma=0,\pm 1$ for the regimes II, I, III respectively, and
\be
\label{appbdef}B=2^{3/2+\Delta}\left(\frac{\Delta}{1-2\Delta}\right)^2\log\left(\frac{4\gamma r_h^2}{e}\right)^2.
\ee
In the outer region, we get
\be
\label{metricfarapp}ds^2=-(r^2-\tilde{r}_h^2)dt^2+\frac{dr^2}{r^2-\tilde{r}_h^2}+r^2d\phi^2+\frac{16\gamma\Delta^2r_h^2U_0}{r^2-r_h^2}dtdr,
\ee
with $\tilde{r}_h\equiv r_h(1-\gamma U_0)$ as before. This has the same form as the far region metric (\ref{metricfaru}-\ref{metricfaruv}), only with a different coefficient in front of the $dtdr$ term, which we expect: far away from the throat, any wormhole will look as a slightly perturbed black hole, no matter what the exact model of the wormhole opening. The Klein-Gordon equation in the near region can be solved at leading order in $\gamma$:
\bea
\nonumber &&\Phi^\mathrm{throat}(t,\rho,\phi;\omega,\ell)=
\rho^{-\frac{1}{2}-\Delta}e^{4\gamma\frac{(\Delta-1/2)^2}{\Delta}r_ht-\imath\frac{\omega\gamma Be^{r_ht}\mathrm{sgn}(t-t_0)}{\Delta\rho^{\Delta+5/2}}}\times\\
&&\times e^{-\imath\omega t+\imath\ell\phi}
{}_1F_1\left(\frac{2-\Delta}{\Delta},\frac{1-\Delta}{\Delta},\frac{2\sqrt{\omega^2-\ell^2} Be^{(2\Delta-1)r_ht}\rho^{-\Delta}}{\Delta}\right).~~
\eea
We have not explicitly written the normalization constant in front. In the far region, we find
\bea
\nonumber&&\Phi^{\mathrm{out}}(t,r,\phi;\omega,\ell)=\left(1+2\imath\gamma U_0\frac{\omega\mathrm{sgn}(t-t_0)}{\sqrt{\omega^2-\tilde{r}_h^2/r^2}}\right)
\exp\left[\imath\gamma\omega U_0\mathrm{sgn}(t-t_0)\left(\frac{r}{r^2-r_h^2}-\arctan\frac{r}{r_h}\right)\right]\times\\
\nonumber&&\times e^{-\imath\omega t+\imath\ell\phi}r^{\imath\tilde{\omega}-\Delta}(r^2-\tilde{r}_h^2)^{-\imath\tilde{\omega}/2}
\label{kgoutapp}{}_2F_1\left(\frac{\imath\tilde{\ell}-\imath\tilde{\omega}+\Delta}{2},\frac{-\imath\tilde{\ell}-\imath\tilde{\omega}+\Delta}{2},\Delta;\frac{r^2}{\tilde{r}_h^2}\right),
\eea
so the outer region solution is of the same form as for the other case ($\Delta>1/2$).

\section{Derivation of the bulk-to-boundary propagator}\label{appb}

In order to perform the sum (\ref{gsum}), we first put $\phi'=0$ as the system is homogenous in the angle $\phi$. This restricts the sum over $\ell$ to $\ell=0$. Second, we expand in $\gamma$ as usual. Finally, for the inner region solutions (\ref{kgthroat}) we will freely change between the coordinates $\rho$ and $r$, with help of (\ref{rho}). By definition, the modes $\Phi^{\mathrm{out}}$ behave well in the interior for any $\omega$. In order to behave well in the whole space (including the boundary), we impose the quantization condition:
\be
\label{hyperquant}\omega=\Delta+\ell+2n\mapsto\Delta+2n,
\ee
taking into account that $\ell=0$; here, $n$ is a non-negative integer. This yields the sum
\bea
\nonumber &&G(r,r';t,t';\phi,\phi')=e^{-\imath\Delta\left(t-t'\right)}\left(\frac{r^2-\tilde{r}_h^2}{r}\right)^\Delta\frac{\gamma\tilde{r}_h}{r-\tilde{r}_h}\times\\
\label{gsum1}&&\times\sum_n e^{-2\imath n\left(t-t'\right)}P_n^{(0,\Delta-1)}\left(r^2-\tilde{r}^2_h\right)K_{\sqrt{1+m^2}}\left(\left(\Delta+2n\right)\frac{r-\tilde{r}_h}{\gamma\tilde{r}_h}\right),
\eea
where $P^{(0,\Delta-1)}$ is the Jacobi polynomial of order $(0,\Delta-1)$, which is obtained from the hypergeometric function satisfying (\ref{hyperquant}). Now we use the identity
\be
K_\nu(x)=e^{-\frac{\imath\pi\nu}{2}}\lim_{N\to\infty}\left(\frac{x}{2N}\right)^\nu P_N^{(\nu,b)}\left(\cos\frac{x}{N}\right)
\ee
to transform (\ref{gsum1}) into a sum of the Jacobi polynomial products as:
\bea
\nonumber &&G(r,r';t,t';\phi,\phi')=e^{-\imath\Delta\left(t-t'\right)-\frac{\imath\pi\sqrt{1+m^2}}{2}}\left(\frac{r^2-\tilde{r}_h^2}{r}\right)^\Delta\frac{\gamma\tilde{r}_h}{r-\tilde{r}_h}\lim_{N\to\infty}\frac{1}{N^{\sqrt{1+m^2}}}\times\\
&&\times\sum_ne^{-2\imath n\left(t-t'\right)}
P_n^{(0,\Delta-1)}\left(r^2-\tilde{r}^2_h\right)\left(\frac{(\Delta+2n)(r-\tilde{r}_h)}{\gamma\tilde{r}_h}\right)^{2\sqrt{1+m^2}}P_N^{(\sqrt{1+m^2},\Delta-1)}\left(\cos\frac{2(r-\tilde{r}_h)}{\gamma\tilde{r}_h}\right).~~~~~~~~~~
\eea
Taking the limit $N\to\infty$ leaves us with a single sum:
\bea
\nonumber &&G(r,r';t,t';\phi,\phi')=\frac{2^\nu}{\sqrt{\pi}}e^{-\imath\Delta\left(t-t'\right)-\frac{\imath\pi\sqrt{1+m^2}}{2}}
\left(\frac{r^2-\tilde{r}_h^2}{r}\right)^\Delta\frac{\gamma\tilde{r}_h}{r-\tilde{r}_h}\left(\frac{r-\tilde{r}_h}{\gamma\tilde{r}_h}\right)^{\sqrt{1+m^2}}\times\\
&&\times\sum_ne^{-2\imath n\left(t-t'\right)}P_n^{(0,\Delta-1)}\left(r^2-\tilde{r}^2_h\right)\left(\frac{(\Delta+2n)\gamma\tilde{r}_h}{r-\tilde{r}_h}\right)^{2\sqrt{1+m^2}}.
\eea
Now we can exploit the Jacobi identities from \cite{propbtz1} to perform the sum with the single remaining Jacobi polynomial. From now on we drop the constant factors as they are not important for our purposes. The result is
\bea
\nonumber &&G(r,r';t,t';\phi,\phi')=
\left[1-\frac{\frac{\gamma (r-\tilde{r}_h)}{\tilde{r}_h}}{\gamma^2+\left(\frac{r-\tilde{r}_h}{\tilde{r}_h}\right)^2\left[e^{\tilde{r}_h(t-t')}\frac{r-\tilde{r}_h}{r+\tilde{r}_h}-e^{-\tilde{r}_h(t-t')}\frac{r-\tilde{r}_h}{r+\tilde{r}_h}+\cosh\left(\tilde{r}_h\left(\phi-\phi'\right)\right)\right]^2}\right]^\Delta\times\\
\label{gbnddirac}&&\times e^{\frac{\gamma U_0}{4}}\left[e^{\tilde{r}_h(t-t')}\frac{r-\tilde{r}_h}{r+\tilde{r}_h}-e^{-\tilde{r}_h(t-t')}\frac{r-\tilde{r}_h}{r+\tilde{r}_h}+\cosh\left(\tilde{r}_h\left(\phi-\phi'\right)\right)\right]^{-\Delta}.
\eea
In the above we have also restored the dependence on $\phi-\phi'$. The remaining step is to apply the relation between $G$ and $K$ (cited in the main text) and to change the coordinates $(t,r)$ to $(U,V)$. This results in Eq.~(\ref{kbnddirac}).

\section{Slow wormhole}\label{appc}

Now we consider the wormhole model which is the opposite case from the fast wormhole -- when the perturbation is turned on for a very long time, starting at very early time $U_0$ and ending at very late time $U_f$. It is logical to call this regime the slow wormhole. The stress tensor for the slow wormhole was obtained already in \cite{wormjaff} by expanding around $U\to\infty$. The stress-energy tensor and the metric correction are
\bea
\label{slowtuu}T_{UU}(U)&=&4\gamma\Delta^2U^{-2\Delta-2}\log U\log\frac{U_f}{U_0}\\
\label{unitstepwhsol}h(U,V)&=&\frac{8\Delta^2}{(1-2\Delta)^2}\log\frac{U_f}{U_0}\frac{1-UV}{1+UV}\frac{1+\left(1-2\Delta\right)\log U}{U^{2\Delta+1}}.
\eea
Note that the strict $U_0\to 0,U_f\to\infty$ limit is not consistent: we cannot reach the eternal wormhole solution by starting from the finite-time double-trace deformation; that would be inconsistent anyway as the coordinates $(U,V)$ would not be well-defined. But we can study the long-time regime which is enough for our purposes. Therefore, we adopt $U_0=1/U_f\to 0$ (so $U_0$ is far in the past and $U_f$ far in the future) and expand in $U_0$ small.\footnote{Remember that $U_0\to 0$ means $t_0\to-\infty$ and $U_0\to\infty$ corresponds to $t_0\to\infty$.}

We now follow the same path as we did for the fast wormhole in the main text. Converting the $U,V$ coordinates to the $t,\rho$ coordinates and dividing the geometry again between the inner and outer region, we find the inner region metric as
\be
\label{metricnearslow}ds^2=(1+\gamma^2r_h^2\rho^2)(-dt^2+d\phi^2)+\frac{d\rho^2}{1+\gamma^2r_h^2\rho^2}+
\frac{\sigma}{2}\log\left(\gamma\rho\right)\frac{2\gamma(\Delta-1)e^{(2\Delta_1-1)r_ht}\rho^{1-\Delta}}{1+\gamma^2r_h^2\rho^2}dtd\rho.
\ee
In the outer region it reads
\be
\label{metricfarslow}ds^2=-(r^2-\tilde{r}_h^2)dt^2+\frac{dr^2}{r^2-\tilde{r}_h^2}+r^2d\phi^2+\frac{8\gamma\log\frac{U_f}{U_0}\Delta^2r_h^2U_0}{(1-2\Delta^2)\left(r^2-r_h^2\right)}dtdr.
\ee

\subsection{Global structure of the geodesics}

\begin{figure}
(a)\includegraphics[width=.28\textwidth]{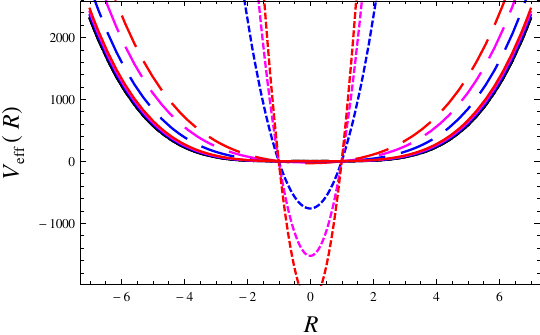}
(b)\includegraphics[width=.28\textwidth]{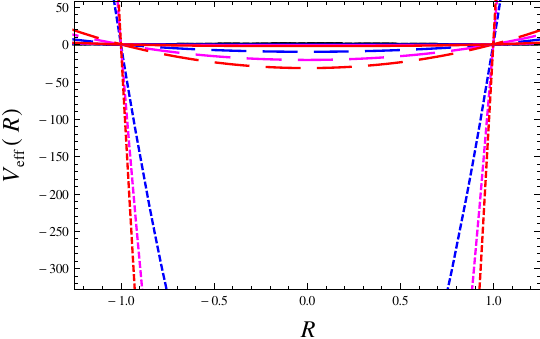}
(c)\includegraphics[width=.28\textwidth]{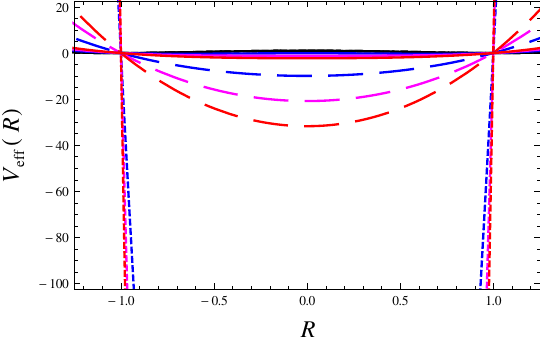}
\caption{The effective potential for a spherical shell (effectively a particle with $L_z=0$) in the background of a fast wormhole with $\Delta=3/4$ (solid lines), fast wormhole with $\Delta=1/4$ (dashed lines) and slow wormhole with $\Delta=1/4$ (dotted lines), each for $\gamma=0,0.1,0.2,0.3$ (black, blue, magenta red); fast wormholes have $U_0=0.5$ and the slow wormhole has $U_0=0.1$. For $\gamma=0$ all three geometries reduce to a BTZ black hole at temperature $T=1/4\pi$, with flat potential on the horizon. For nonzero coupling, a well develops for small $R$, but the well is shallow for fast wormholes and quite deep for slow wormholes, implying many bound states which however are not infinitely stable -- their lifetime grows with the height of the maxima of $V_\mathrm{eff}$ at intermediate $r$ values. The panels (a, b, c) zoom in successively in the small $r$ region, to show clearly both the shallow wells and the deep wells. For very large $r$ (not in the range of the plots) all potential wells acquire the universal form imposed by the AdS asymptotics.}
\label{figveff}
\end{figure}

One important effect of the near-eternal nature of the wormhole, that can be significant already at leading order in $\gamma$, is the possibility of an orbit which goes back and forth multiple times, contributing to the backreaction whenever it passes near $UV=0$. This is a difficult topic. Geodesics in the wormhole geometry are nonintegrable, since the only conserved quantity is the angular momentum (which is trivial anyway for a spherical wave); energy is not conserved because the geometry changes explicitly at $U_0$. We will give a few numerical examples to demonstrate that multiple windings are an exception in fast wormholes but almost a rule in slow wormholes. This is logical: a fast wormhole only has the left-right coupling for an instant, thus most orbits do not have time to go back and forth; for a slow wormhole there is ample time. A convenient way to see this is to start from the Lagrangian in the Schwarzschild coordinates
and formulate the effective potential $V_{\mathrm{eff}}$:
\bea
\label{lageik}&&L=\frac{\dot{R}^2}{r_h^2-R^2}+r_h^2-R^2-\gamma r_h^2\frac{\left(\dot{R}^2-r_h^2+R^2\right)\left(e^{2r_h\tau}h-e^{-2r_h\tau}\tilde{h}\right)}{(r_h^2-R^2)(r_h+R)^2}\\
\label{veffeik}&&V_{\mathrm{eff}}=-\left(r_h^2-R^2\right)^2+\gamma r_h^2\frac{\left(R^2-r_h^2\right)\left(e^{2r_h\tau}h-e^{-2r_h\tau}\tilde{h}\right)}{(r_h+R)^2}
\eea
It is understood that $R=R(\tau)$, and the angular momentum $L_z$ is put to zero. Now we can plot the effective potential as a function of the wormhole coupling $\gamma$ and the wormhole creation time $U_0$ (inserting of course the expressions for $h,\tilde{h}$ for a chosen wormhole model). Fig.~\ref{figveff} shows the effective potential for a fast wormhole with $\Delta=3/4$, a fast wormhole with $\Delta=1/4$ and a slow wormhole with $\Delta=1/4$, for a range of $\gamma$ and $U_0$ values. Two effects are obvious (1) the effective potential, flat at the horizon for a BH ($\gamma=0$), develops a well whose depth with the left-right coupling $\gamma$ and the lifetime of the wormhole (2) while fast wormholes never develop a very deep well, the depths grows drastically in the slow wormhole approximation. We could try to study the high-lying bound states in deep wells within WKB formalism and count them analytically, but we postpone such detailed investigations of the throat dynamics for later work. For now we are content to conclude that \emph{slow wormholes have many bound states\footnote{These bound states are not infinitely
long-living because the well in the center is of finite depth, meaning that the particle will eventually escape.} which become denser and denser as they approach the top of the well}. Therefore, orbits can likely spend a long time inside the throat.

\begin{figure}
(a)\includegraphics[width=.28\textwidth]{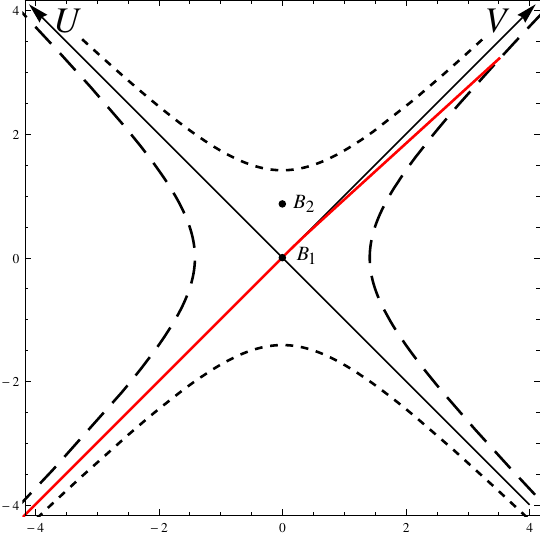}
(b)\includegraphics[width=.28\textwidth]{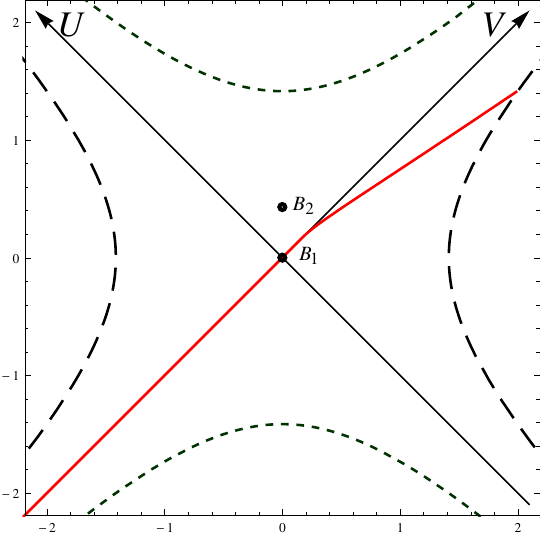}
(c)\includegraphics[width=.28\textwidth]{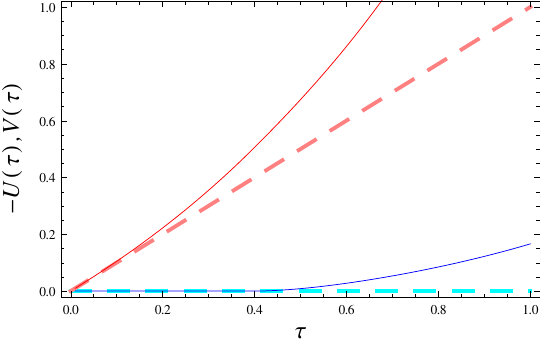}
\caption{A typical orbit (red) in a fast wormhole background for $\Delta=1/100,\gamma=0.1$ (a) and for $\Delta=1/4,\gamma=0.5$ (b), in Kruskal coordinates $(U,V)$. The $U$- and $V$-axis form the horizon for $\gamma=0$ (black hole), the singularity is reached for $UV=1$ (dotted black lines) and the AdS boundaries are at $UV=-1$ (dashed black lines); the original bifurcation surface is at $B_1$ and the point of left/right future horizon crossing is $B_2$. In (c) we show the coordinates $-U(\tau)$ (blue) and $V(\tau)$ (red) for $\Delta=1/2$, and $\gamma=0$ (BH, dashed lines) or $\gamma=0.1$ (wormhole, solid lines). The divergence from the BH geodesic $U=0,V=\tau$ is relatively small, justifying a perturbative treatment.}
\label{figgeod}
\end{figure}

Let us also look at the orbits. In Fig.~\ref{figgeod} we show a geodesic in a fast wormhole geometry and for reference also for a BTZ black hole. Of course, a wormhole replaces the horizon by a throat so instead of falling into the singularity the particle goes to the other AdS boundary; this may repeat several times if there are windings. But for small $UV$, the two orbits are remarkably close, as we see also by looking at the coordinates in proper time $U(\tau), V(\tau)$: they only diverge from the BTZ BH in a perturbative way. This suggests a practical method for computing the backreaction: we focus on the region where the boost is maximized, i.e. the minimal $UV$ part of the orbit, where its stress tensor can be obtained as a series expansion in $\gamma$ around the stress tensor for a particle/spherical wave in the BH geometry. The redshift is now finite everywhere, but it has a sharp maximum in the far IR region which is thus still the crucial one for scattering. 

\subsection{The backreaction of the perturbation and the eikonal phase}

The general results (\ref{shockshift}-\ref{tmunutot}) for the backreaction still hold. Inserting the metric (\ref{unitstepwhsol}) into the expressions for the backreaction, we find that $\delta g_{\phi\phi}$ again does not change compared to (\ref{diracphph}), and for the remaining functions we get
\bea
\label{slowwhc}c(U,V)&=&2p_0-2\gamma^2p_0U_0^2(\epsilon V)^{-\Delta-1}(\log\epsilon V)\\
\label{slowuv}\delta g_{UV}&=&\frac{\gamma p_0}{(1+UV)^2}.
\eea
Inserting this into the action (\ref{linact}), we get
\be
\label{linactunit0}S_c^{(\mathrm{slow})}=p_0q_0-2\gamma U_0^{-2-3\Delta}\left(4p_0q_0U_0\log U_0+2\left(p_0^2+q_0^2\right)\log U_0+O\left(U_0^2\right)\right).
\ee
Remember that the slow WH approximation consists in expanding in $U_0$ small and taking only the leading term. In (\ref{linactunit0}), the BH contribution $p_0q_0$ (the first term) is suppressed by a factor of $U_0^{3\Delta+2}$ when $U_0\to 0$ (ignoring the subleading logarithms), and the second term, also proportional to $p_0q_0$, is suppressed by a factor of $U_0$. Therefore, the dominant contribution is just the third term:
\be
\label{linactunit}S_c^{(\mathrm{slow})}=-4\gamma U_0^{-2-3\Delta}\log U_0\left(p_0^2+q_0^2\right).
\ee
Here the qualitative difference from BH scrambling is obvious: fast scrambling is the consequence of the on-shell action being equal to the center-of-mass momentum squared \cite{scramble2}, which equals $p_0q_0$. Now this term is suppressed by $U_0$ small, and the scrambling is determined mainly by the asymmetric contributions $p_0^2$ and $q_0^2$: we can already guess something will change drammatically compared to the black hole scrambling.

\subsection{Lyapunov spectra for slow wormholes}

In a slow wormhole the chaos apparently becomes exponentially weak or nonexistent. We have seen in (\ref{linactunit0}-\ref{linactunit}) that at leading order in $U_0$ (which is roughly the expansion in large $-t_0$, with $t_0<0$) the eikonal phase factorizes into $\sim\exp(-p_0^2-q_0^2)$. If this factorization were exact, OTOC would behave exactly as TOC at long times, i.e. as a simple product of expectation values. However, higher order terms in the classical action and the products of the propagators couple $p_0$ and $q_0$. After integrating out the coordinate dependence, we arrive at:
\bea
\nonumber &&D^{(\mathrm{slow})}(t,0)=\int d\tilde{p}\int d\tilde{q}\tilde{p}^{2\Delta_1-1}\tilde{q}^{2\Delta_2-1}(\tilde{p}+\tilde{q})^{\Delta_1+\Delta_2-2}\times\\
\label{otocintslow}&&\times\exp\left[\left(1-2\left(2+3\Delta_1\right)^{-1}\gamma U_0^{-1-3\Delta_1}\right)\left(\tilde{p}^2+\tilde{q}^2\right)\right].
\eea
This integral is harder to solve than the previous ones, but we can still do the saddle-point integration over $\tilde{q}$ (obtaining generalized hypergeometric functions as a result), and then the $\tilde{p}$ integral is doable through a series expansion assuming the main contribution comes from small $\tilde{p}$ -- this is just an assumption that we cannot justify rigorously. When everything is over, we get:
\bea
\nonumber D^{(\mathrm{slow})}(t,0)&\sim&\left[1-\exp\left(\left(2+3\Delta_1-2\gamma U_0^{-1-3\Delta_1}\right)^2e^{-(2+3\Delta_1)/2\pi T}t\right)\right]^{-2\Delta_2}\Rightarrow\\
\label{otocslow}&\Rightarrow&\lambda=\left(2+3\Delta_1-2\gamma  U_0^{-1-3\Delta_1}\right)^2e^{-(2+3\Delta_1)/2\pi T}~\sim~e^{-(2+3\Delta_1)/2\pi T}~~~~~~~~~~~~
\eea
Therefore, even though the growth exponent is positive, it is exponentially small. We do not have a good analytical understanding of this regime as the small-$p$ expansion of the integrand is of questionable validity.

\begin{figure}
(a)\includegraphics[width=.4\textwidth]{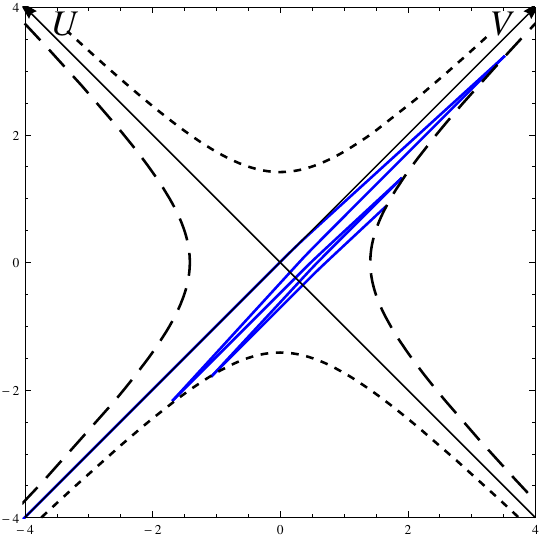}~~~~
(b)\includegraphics[width=.4\textwidth]{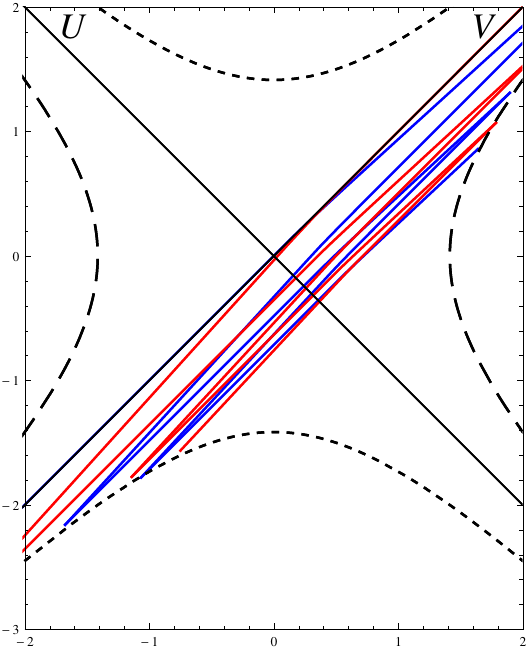}
\caption{(a) An orbit in a slow wormhole background (here for $\Delta=1,\gamma=0.5$) can go back and forth many times, which makes analysis difficult. But if we consider at the same time the dynamics of the in-wave (blue) and the out-wave (b) we see that the two orbits are almost parallel most of the time and there is hardly any scattering going on. This provides an intuitive explanation why the Lyapunov exponent is very small.}
\label{figslow}
\end{figure}

There is another way of looking at the dynamics of OTOC in the slow wormhole background which is possibly more enlightening: we can forgo the momentum space calculations and look at the perturbations in the coordinate space.\footnote{We thank to Zhengbin Yang for suggesting this viewpoint.} When the wormhole is almost eternal, both waves can go back and forth many times but the important point is not only the shape of each orbit separately but also how much they scatter. The answer is -- very little, if the black hole horizon vanishes far in the past and reforms in far future. This can be seen in Fig.~\ref{figslow}. Not only does an orbit bump back and forth many times, leading to recurrences, but also the IN wave and the OUT wave are almost parallel in the interior so their scattering cross section is very small. This is of course just a rephrasing of the finding (\ref{linactunit}) that the classical action contains no center-of-mass term proportional to $pq$ at leading order, but here it is seen directly from the kinematics.

\end{document}